\documentclass{aa}  

\usepackage{graphicx}
\usepackage{txfonts}
\usepackage{natbib}
\usepackage{makecell}
\bibpunct{(}{)}{;}{a}{}{,}

\usepackage{hyperref} 
\begin{document}

   \title{A multiwavelength overview of the giant spiral UGC~2885}

   \author{Matheus C. Carvalho
   %https://orcid.org/0000-0002-5568-6965
          \inst{1} %\orcidicon{0000-0002-5568-6965}
          \and
          Bavithra Naguleswaran\inst{1} %\orcidicon{0009-0005-5255-8238}
          \and
          Pauline Barmby\inst{1, 2} %\orcidicon{0000-0003-2767-0090}
          %https://orcid.org/0000-0003-2767-0090
          \and
          Mark Gorski\inst{3, 4} %\orcidicon{0000-0001-9300-354X}
          %https://orcid.org/0000-0001-9300-354X
          \and
          Sabine K\"oenig\inst{4} %\orcidicon{0000-0001-6174-8467}
          %https://orcid.org/0000-0001-6174-8467
          \and
          Benne Holwerda\inst{5} %\orcidicon{0000-0002-4884-6756}
          %https://orcid.org/0000-0002-4884-6756
          \and
           Jason Young\inst{6, 7} %\orcidicon{0000-0002-5830-9233}
          }

\institute{Department of Physics \& Astronomy, University of Western Ontario, 1151 Richmond Street, London, Ontario, Canada
\and
Institute for Earth \& Space Exploration, University of Western Ontario, 1151 Richmond Street, London, Ontario, Canada
\and
Center for Interdisciplinary Exploration and Research in Astrophysics (CIERA) and Department of Physics and Astronomy, Northwestern University, Evanston, IL 60208, USA
\and
Department of Space, Earth and Environment, Chalmers University of Technology, Onsala Space Observatory, 439 92 Onsala, Sweden
\and
Department of Physics \& Astronomy, University of Louisville, Natural Science Building 102, Louisville, KY 40292, USA
\and
Department of Astronomy, University of Massachusetts, Amherst, MA 01003, USA
\and
The SETI Institute, Mountain View, California 94043, USA
}

\date{Received --, --; accepted --, --}

\abstract
   {UGC~2885 ($z = 0.01935$) is one of the largest and most massive galaxies in the local Universe, yet its undisturbed spiral structure is unexpected for such an object and unpredicted in cosmological simulations. Understanding the detailed properties of extreme systems such as UGC 2885 can provide insight on the limits of scaling relations and  physical processes driving galaxy evolution.}
   {Our goal is to understand whether UGC~2885 has followed a similar evolutionary path to other high-mass galaxies by examining its place on the fundamental metallicity relation and the star-forming main sequence.}
   {We present new observations of UGC~2885 with the Canada-France-Hawaii, and Institut de radioastronomie millimétrique 30-m, telescopes. These novel data are used to respectively calculate metallicity and molecular hydrogen mass values. We estimate stellar mass (M$_{\star}$) and star formation rate (SFR) based on mid-infrared observations with the Wide-field Infrared Survey Explorer.}
   {We find global metallicities $Z = 9.28$, 9.08 and 8.74 at the 25~kpc ellipsoid from N2O2, R23 and O3N2 indices, respectively. This puts UGC~2885 at the high end of the galaxy metallicity distribution. The molecular hydrogen  mass is calculated as M$_{\rm H_{2}}=1.89 \pm 0.24 \times 10^{11}$~M$_{\odot}$, the SFR as $1.63 \pm 0.72$~M$_{\odot}$~yr$^{-1}$ and the stellar mass as $4.83 \pm 1.52 \times 10^{11}$~M$_{\odot}$, which gives a star formation efficiency (SFE = ${\rm SFR}/\rm M_{\rm H_{2}}$) of $8.67 \pm 4.20 \times 10^{12}$~yr$^{-1}$. This indicates that UGC~2885 has an extremely high molecular gas content when compared to known samples of star forming galaxies ($\sim100$ times more) and a relatively low SFR for its current gas content.}
   {We conclude that UGC~2885 has gone through cycles of star formation periods, which increased its stellar mass and metallicity to its current state. The mechanisms that are fueling the current molecular gas reservoir and keeping the galaxy from producing stars remain uncertain. We discuss the possibility that a molecular bar is quenching star forming activity.}


   \maketitle
%
%-------------------------------------------------------------------

\section{Introduction}

The galaxy UGC~2885, or Rubin's Galaxy, is notable in the local Universe  ($z<0.1$) for being the largest and most massive spiral \citep{rubin1980rotational, romanishin1983very}. 
UGC~2885 has four defined arms and traces of star formation activity along them \citep{hunter2013star}.
With an estimated radius of $122$~kpc and a total mass of M$_{\rm tot} = 1.5 \times 10^{12}\; \rm M_{\odot}$ \citep{rubin1980rotational}, this object challenges the understanding of galactic formation and evolution:
cosmological simulations \citep[e.g.][]{vogels2014,snyder2015} do not predict the existence of such galaxies.
The higher end of the mass distribution of galaxies is thought to have evolved by the accretion of gas by merger events with companion or generally nearby systems \citep{lacey1993merger,kerevs2005galaxies}. This might not be the case for UGC~2885, as it lives in an isolated environment, is essentially bulge free and shows no sign of perturbed morphology.

Another consequence of growing galaxies by mergers is the presence of an active galactic nucleus (AGN) since mergers diminish angular momentum and contribute to black hole accretion \citep{dietrich2018agn, gao2020mergers}. 
\citet{holwerda2021predicting} compared a predicted UGC 2885 spectrum predicted by machine learning to 
spectra obtained from optical observations with VIRUS-P \citep[Visible Integral-field Replicable Unit Spectrograph – Prototype;][]{2008SPIE.7014E..70H}
and MMT/Binospec.
They found good qualitative agreement and concluded that both predicted and observed line ratios 
(\ion{O}{III}/H$\beta$ versus \ion{N}{II}/H$\alpha$ and \ion{O}{III}/H$\beta$ versus \ion{S}{II}/H$\alpha$)
were consistent with AGN activity. 
The mid-infrared colour of UGC 2885's central region was not sufficiently red to identify it as an AGN,
and \citet{holwerda2021predicting} suggested that a mid-infrared AGN signal is likely diluted by the large point spread function (PSF). 

Although massive, UGC~2885 does not meet the criteria to be considered a 'super spiral' \citep{ogle2016superluminous}. Those galaxies are generally found in richer environments and have higher star formation rates (SFR) than those reported for Rubin's Galaxy: $5 - 70$~M$_{\odot}$~yr$^{-1}$ for super spirals \citep{ogle2016superluminous} versus $2.47$~M$_{\odot}$~yr$^{-1}$ for UGC~2885 \citep{hunter2013star}.% 
\footnote{When literature measurements are quoted without uncertainties, no uncertainty values were given in the previous work.}
Another analogue classification is the group of giant low surface brightness disc galaxies \citep[gLSB; ][]{saburova2021observational}. These objects are remarkably isolated, contain large reservoirs of gas and their star formation is extremely low \citep[mean SFR of $0.88$~M$_{\odot}$~yr$^{-1}$; ][] {du2023star}. UGC~2885 has a higher SFR and a typical surface brightness disc than gLSBs \citep{holwerda2021predicting}.

Previous studies have acknowledged that UGC~2885 lies on the extreme ends of the mass and size distribution for local galaxies. $21 \; \rm cm$ line observations using the Westerbork Synthesis Radio Telescope (WSRT) probed the rotation curve derived from neutral hydrogen (\ion{H}{I}) emission of this object, corroborating the exceptional baryonic mass and morphological symmetry \citep{lewis1985hi, roelfsema1985radio}. \citet{canzian1993spiral} obtained H$\alpha$ imaging data of UGC~2885 using the Isaac Newton Telescope  and found that underlying density waves may contribute to the stability of this galaxy's disc, possibly driven by its enormous mass. Expanding on the same optical line with the Kitt Peak National Observatory (KPNO), \citet{hunter2013star} identified ionised hydrogen (\ion{H}{ii}) regions across $5.6$ disc scale lengths ($\sim70 \; \rm kpc$) as well as detached star forming regions at the end of the galactic arms. 

In this paper, we aim to provide an overview of some of Rubin Galaxy's properties across a range of observations. By representing a system at the upper bounds of mass and size in the nearby Universe, UGC~2885 could be an important tool to better understand galaxy formation and how certain properties can evolve in an isolated environment. Although it has been previously studied in the radio and optical regimes, we will extend   results by diving into specific lines within those bands. In the mm regime, we introduce CO($1-0$) (rest frequency 115.27 GHz) line observations, a probe of molecular hydrogen (H$_{2}$), and in the optical we study the \ion{O}{II}$\lambda3727$, H$\beta \lambda 4861$,\ion{O}{III}$4959+5007$, H$\alpha \lambda 6563$, \ion{N}{II}$6548+6583$, \ion{S}{II}$6717+6731$ emission lines, analysing their line ratios to provide insight on metallicity
and, although not the major focus of this work, nuclear activity.
We aim to answer the following questions: How does UGC~2885 compare to other massive spiral galaxies on the stellar mass-star formation rate plane? Does it seem to evolve the same way as normal-sized spirals? What are the possible evolution scenarios that made this galaxy grow to its great mass without losing spiral structure and active star formation?

This paper is organised as follows: Section \ref{sec:obs} discusses the archival and novel observations of UGC~2885, as well as how each specific dataset was processed and analysed. 
Sections~\ref{sec:met}--\ref{sec:masses} discuss the estimation of physical properties from the observational data.
In Section \ref{sec:results}, we explore the results of our analysis, focusing specifically on the new information found---molecular gas content and stellar population---and their relation to the current understanding of this object's formation and evolution. Section \ref{sec:conc} summarises the findings and points to future work.

\section{Observations}\label{sec:obs}

\begin{table}
    \caption{Basic properties of UGC~2885}
        \begin{tabular}{lll}
        \hline \hline
        Parameter                    &  & Reference \\ \hline
        Right Ascension ($\deg$)        & $ 58.268000$ & (1)\\
        Declination ($\deg$)            & $ 35.591944$ & (1)\\
        Inclination ($\deg$)            & $65$ & (2)\\
        Redshift & $0.01935$ & (3) \\
        Heliocentric Velocity (km~s$^{-1}$) & $5801 \pm 3$ & (3)\\
        Distance (Mpc)               & $ 84.34 \pm 5.68 $ & (3)\\
        Half-light Radius (V-band, kpc)   & $22.2$ & (4)\\
        Disc Scale Length (V-band, kpc)   & $12.05 \pm 0.41$ & (4)\\
        Star Formation Rate (M$_{\odot}$~yr$^{-1}$) & $2.47$ & (4)\\
        Morphology                   & SA(rs)c & (5)\\
        E(B-V)                       & 0.206 & (6)\\
        Stellar mass (M$_{\odot}$) & $1.58 \pm 0.73 \times 10^{11}$ &(7)\\
        \hline
        \end{tabular}
        \tablefoot{
          References: (1) \cite{skrutskie2006two}, (2) \cite{rubin1980rotational}, (3) \cite{falco1999updated} - Distance calculated from the heliocentric velocity (assuming H$_{0} = 70$~km~ s$^{-1}$~Mpc$^{-1}$, $\Omega = 0.3$), (4) \cite{hunter2013star}, (5) \cite{de2013third}, (6) \cite{chiang2023corrected}, (7)  \citet{di2023dark}.}
\label{tab:params}
\end{table}

\begin{figure}
    \centering
    \includegraphics[width=\columnwidth]{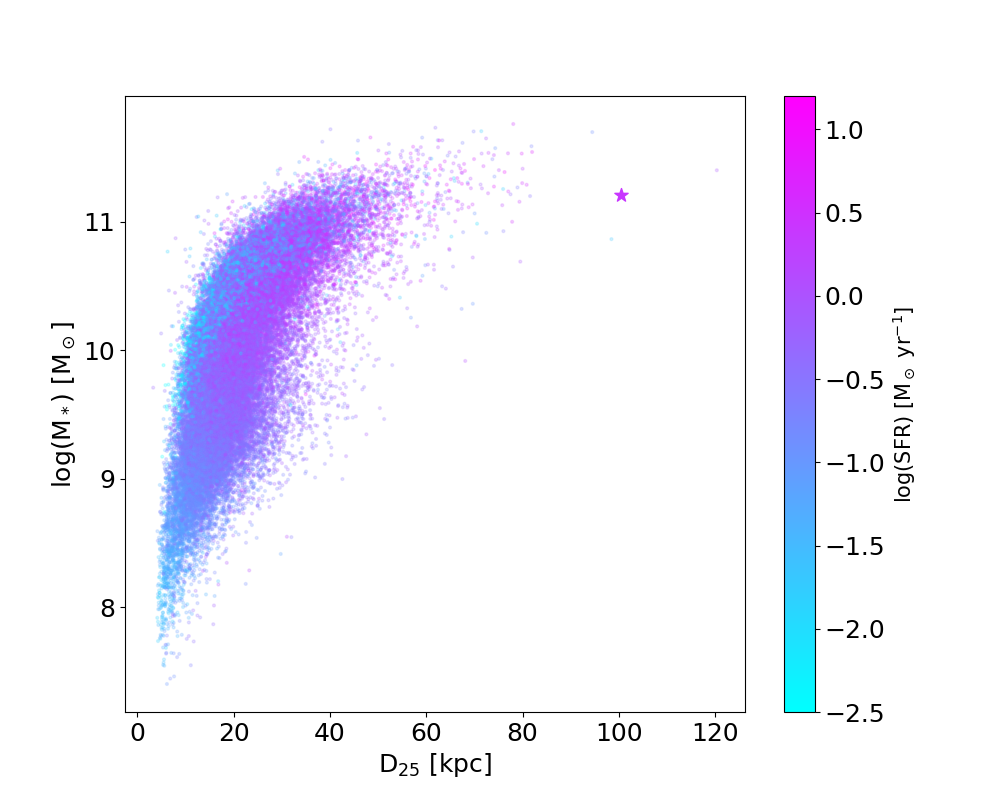}
    \caption{Size-stellar mass relation for local  star-forming galaxies in the
    GALEX-SDSS-WISE Legacy Catalog \citep{GSWLC2016} and Siena Galaxy Atlas \citep{sga2023}. Large star indicates the position of UGC~2885, with stellar mass as reported by \citet{di2023dark}, star formation rate as reported by \cite{hunter2013star}, and $D_{25}$ from the HyperLEDA database and the distance given in Table~\ref{tab:params}.}
    \label{fig:size_mstar}
\end{figure}

\begin{figure}
    \centering
    \includegraphics[scale=0.14]{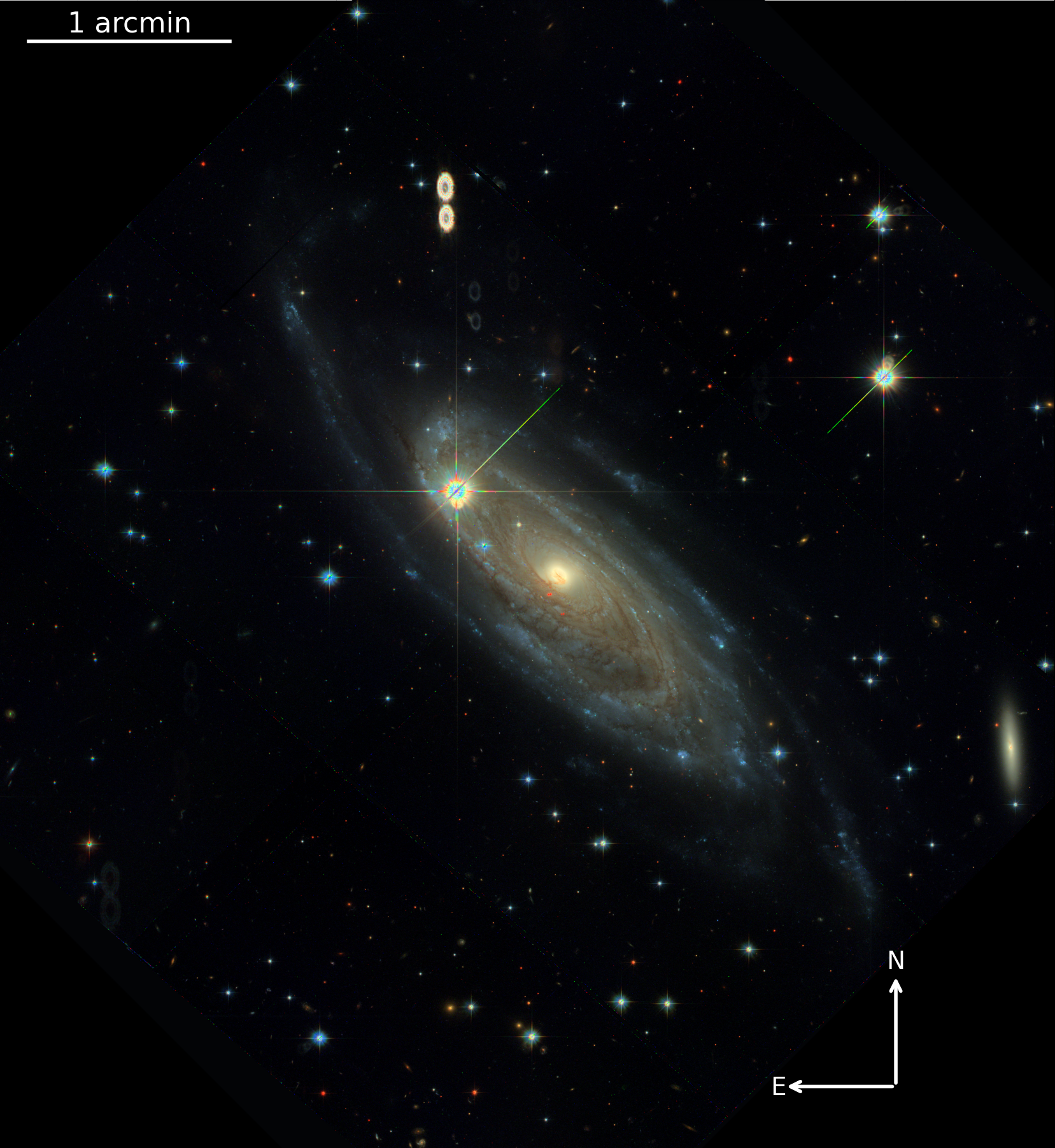}
    \caption{Colour composite of  HST/WFC3 images of UGC~2885 (F814W, F606W and F475W in red, green and blue, respectively). A bright foreground star (HD279085, $V = 10.7$) is inconveniently superimposed on the disc.}
    \label{fig:hstimage}
\end{figure}

Table \ref{tab:params} summarises some of UGC~2885's properties.
To place UGC~2885 in context, Figure~\ref{fig:size_mstar} compares the galaxy's size and stellar mass 
to those of $5.3\times10^4$ local ($z<0.05$) star-forming galaxies in the combination of the
GALEX-SDSS-WISE Legacy Catalog \citep{GSWLC2016} and Siena Galaxy Atlas \citep{sga2023}.
It is one of the largest and most massive galaxies in the local Universe.
For visual reference, Figure \ref{fig:hstimage} shows a colour composite of Hubble Space Telescope (HST) Wide Field Camera 3 (WFC3) images of UGC~2885 (Proposal ID: 15107, P.I.: Holwerda), constructed with the \textsc{reproject} and \textsc{make\_lupton\_rgb} packages of the Astropy library. 
Detailed analysis of the HST images, including surface brightness profile fitting and star cluster 
population analysis will be presented elsewhere (Holwerda et al., in prep.; see also Section~\ref{sec:bar}.)

 Our data consist of a multiwavelength collection of images and datacubes that spans from the optical to the radio domain, mixing both archival and novel observations summarised in Table~\ref{tab:obs}. Luminosities and subsequent properties were calculated assuming a flat cosmology with H$_{0} = 70$~km~s$^{-1}$~Mpc$^{-1}$, $\Omega_{\rm m} = 0.3$ and $\Omega_{\Lambda} = 0.7$.
At the distance of UGC 2885, 1\arcsec corresponds to a physical projected distance of 409 pc; UGC~2885's half-light radius of 22~kpc subtends 0\farcm9.

\subsection{Optical: SITELLE}
\label{sec:sitelle_data}

We observed UGC~2885 with the Canada-France-Hawaii Telescope (CFHT)'s SITELLE
(Spectromètre Imageur à Transformée de Fourier pour l'Etude en Long et en Large de raies d'Emission), an imaging Fourier transform spectrograph specifically designed to detect prominent emission lines in the visible spectrum \citep{drissen_sitelle_2019}. SITELLE provides an $11\arcmin \times 11\arcmin$ field of view (FOV) with a plate scale of 0\farcs32 per pixel. This FOV allows us to achieve a comprehensive single-field view of the galaxy, resulting in datacubes covering three relatively narrow wavelength ranges. Observations were carried out under proposal IDs 19BC13 and 20BC05 with observations performed on 30 Sep 2019, 19 Nov 2020 and 11 Dec 2020. Seeing during the observation dates varied between approximately 1\arcsec and 1\farcs4 full-width at half-maximum (FWHM).

\begin{table*}
    \caption{Observational data}
    \label{tab:obs}
    \begin{tabular}{lllll} 
     \hline \hline
    Band & Wavelength & Spectral resolution & Itime & Spatial resolution\\
    & (frequency)& $R=(\lambda/\Delta \lambda)$& & FWHM \\
        \hline
         \multicolumn{5}{l}{SITELLE:  $11\arcmin \times 11\arcmin$ FOV}\\
         SN1 & 363--386 nm & $R=1000$ & 10802 & 1\farcs2\\ 
        SN2 & 482--513 nm & $R=1000$ & 10797 & 1\farcs2\\
        SN3 & 647--685 nm & $R=5000$ & 14398 & 1\farcs2\\
         \multicolumn{5}{l}{WISE: all-sky}\\
         W1 & 3.4 $\mu$m & $R\approx3.7$ & n/a  & 6\farcs1\\ 
         W2 & 4.6 $\mu$m & $R\approx4.1$ & n/a & 6\farcs8\\
         W3 & 12 $\mu$m & $R\approx1.1$ & n/a & 7\farcs4\\
         W4 & 24 $\mu$m & $R\approx4.8$ & n/a & 12\farcs0\\
         \multicolumn{5}{l}{IRAM:  $5\arcmin \times 1.6\arcmin$ FOV}\\
E0 & 112.29--113.59 GHz & 25 km s$^{-1}$& n/a & 22\farcs5\\
\hline
    \end{tabular}
    \tablefoot{
``Itime" for SITELLE refers to total time to acquire the datacube.}
\end{table*}

We used the CFHT-provided SITELLE data \citep[processed with the ORBS (Outil de Réduction Binoculaire pour SITELLE) pipeline]{martin2012} and carried out further analysis with LUCI \citep{Carter_2021,rhea_luci_2021}.
ORBS-processed deep white-light images were utilised to view details of the galaxy while LUCI was used to produce line flux maps for each strong emission line from the ORBS-procesed datacubes.
We followed the template Python notebooks provided in LUCI (BasicExample, SN1SN2), adopting the suggested parameters for the line shape fitting function of sinc convolved with Gaussian for SN1 and pure Gaussian for SN2 and SN3.
The templates were changed to reflect the spectral resolution used in our observation (Table~\ref{tab:obs}), the galaxy's redshift, and our choice to use $2\times2$ pixel binning.
Binning doubles the signal-to-noise ratio (SNR) and reduces processing time, at a cost of reducing the spatial resolution.
The observations for each filter were taken on separate dates, so the datacubes did not have identical pointing. We adjusted the x and y pixel parameters in the LUCI fit\_cube function, using bright stars as a guide, to ensure alignment between all three datacubes.  
The aligned line flux maps (and corresponding uncertainty maps) generated with LUCI were combined into line ratio maps to measure metallicity (see Section~\ref{sec:met}) and nuclear line ratios (Appendix~\ref{sec:apdx}).

\begin{figure*}
\centering
    \includegraphics[scale=0.4]{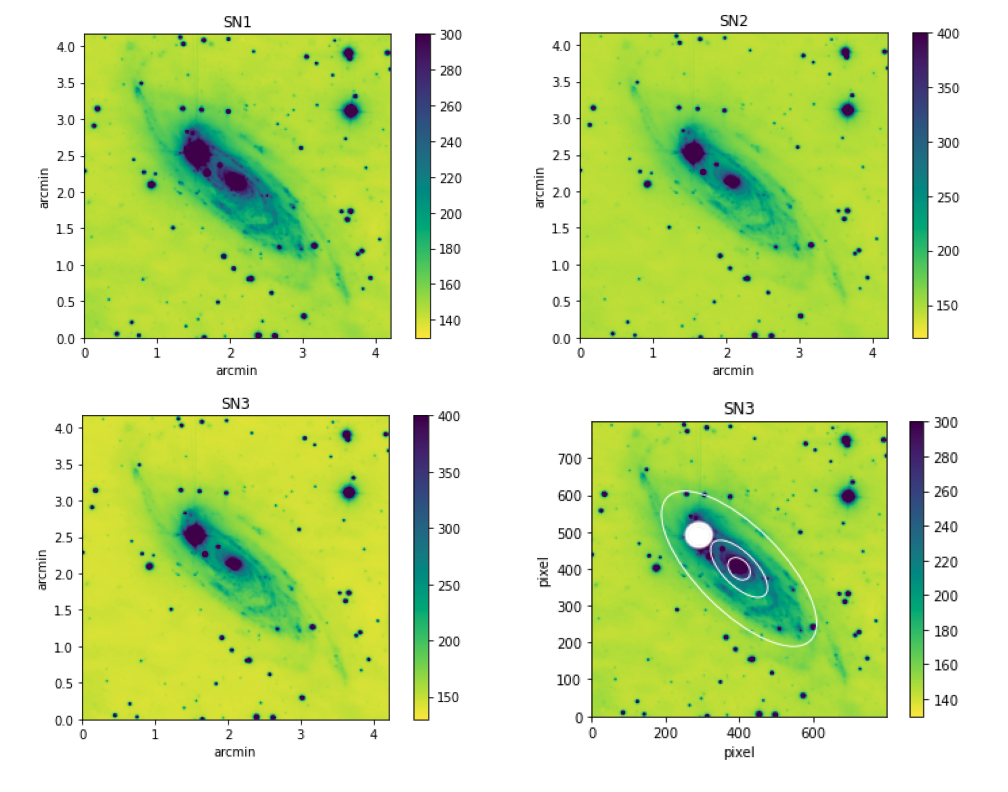}
    \caption{ORBS processed deep white light flux maps in SN1 (363--386~nm), SN2 (482--513~nm) and SN3 (647--685~nm) filters. 
    Bottom right: Illustrates the placement of elliptical apertures used to calculate metallicities and the metallicity gradient. The white circle represents a bright foreground star, which was masked during metallicity calculations.}
    \label{fig:fluxmap}
\end{figure*}

The ORBS processed deep white light images (Figure \ref{fig:fluxmap})
reveal similar galaxy structure to the HST images (Figure \ref{fig:hstimage}), although at about 10 times lower spatial resolution (about 400~pc for SITELLE and 40~pc for HST).
The white light images are sufficient for use in defining three elliptical apertures with which to measure spatial variation in metallicity. The inner aperture focuses on the central bulge while the largest ellipse allows a determination of the global metallicity. We opted not to analyse the outer spiral arms due to their low surface brightness. As Figure \ref{fig:fluxmap} shows, we also masked the bright foreground star that appears in the outer ellipse. 

All flux maps were corrected for foreground dust extinction. We used the \textsc{extinction} package \citep{barbary2016extinction} applying the \textsc{fm07} function, which refers to the \citet{fitzpatrick2007analysis} extinction law, the most recent available with the code. This function assumes a ratio of absolute to selective extinction $R_{\rm V} = A_{\rm V}/E(B-V)= 3.1$ \citep{1989ApJ...345..245C}. Using different extinction laws  did not change the line ratios appreciably.

\subsection{Infrared: WISE}

We use images from the Wide-field Infrared Survey Explorer \citep[WISE; ][]{wright2010wide}, available from the NASA/IPAC Infrared Science Archive\footnote{\url{https://irsa.ipac.caltech.edu/}} to calculate the Star Formation Rate. The WISE telescope, launched in 2009 as a mid-IR ($3.4, 4.6, 12$ and 22~$\mu$m) all-sky survey, has a 40~cm diameter primary mirror with a spatial resolution of $\sim6\arcsec$ for W1, W2 and W3 and $\sim12\arcsec$ for W4. Each of the four bands covers a field of view of $47\arcmin \times 47\arcmin$ and the images are combined into a mosaic that subtends a $1.56^\circ \times 1.56^\circ$ wide field footprint, meaning a co-added image with $4095 \times 4095$ pixels (with a pixel scale of $1\farcs375$~pixel$^{-1}$). For the UGC~2885 observations, bands 1 and 2 have 255 frames co-added to the final footprint and bands 3 and 4 have 151 frames.

The two WISE bands at the longest wavelengths are used as SFR indicators considering that mid-infrared dust emission represents reprocessed ultraviolet (UV) light from newly formed stars. WISE band 3 (W3), centred at 11.6~$\mu$m, is characterised as the Polycylic Aromatic Hydrocarbon (PAH) band, as it is close to PAH features that could trace star forming activity \citep{sandstrom2010spitzer} and band 4 \citep[W4, suggested to be actually centred at 23~$\mu$m after recalibrations][]{brown2014recalibrating} is thought to indicate better SFR measurements considering that it displays the warm dust continuum and it is not affected by emission lines in the low redshift Universe \citep{brown2017calibration}.

Although the AllWISE Source Catalog \citep{cutri2021vizier} contains the magnitudes for each WISE band, we decided to derive those values ourselves. The argument for this is that UGC~2885 is recognised by the catalogue as an extended source and the standard aperture measurements, combined with the co-addition of frames, are developed for point sources as it uses a resampling method based on the telescope's PSF, underestimating the flux of resolved targets. Accordingly, the Wise Explanatory Guide \citep{cutri2012explanatory} recommends a large aperture photometry procedure to correctly estimate the in-band magnitudes.

\begin{figure}
    \centering
    \includegraphics[scale=0.3]{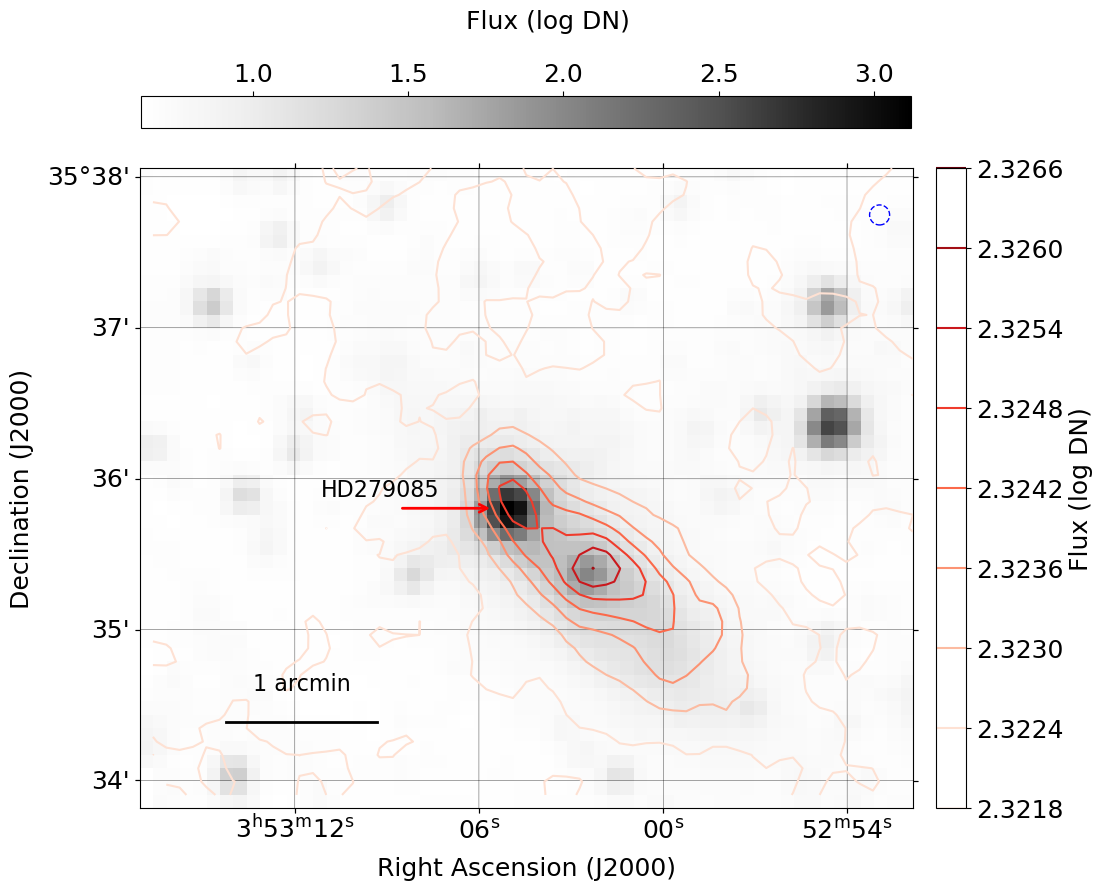}
    \caption{WISE images of UGC~2885 in bands W1 (3.6~$\mu$m; grey scale) and W4 (22~$\mu$m; contours). Blue dashed circle is the W1 beam size (3\farcs98).}
    \label{fig:w1w4}
\end{figure}

Subsequently, we used the SAOImageDS9 \citep{joye2003new} software to perform same-region elliptical aperture photometry on all four WISE bands to obtain global galaxy properties. UGC~2885 has in its field a bright foreground star that dominates the two shorter-wavelength bands (W1 and W2).  Figure \ref{fig:w1w4} shows how the star is brighter than the galaxy in the W1 image but is much fainter in the W4 image. We remove the stellar contribution from the total galactic flux for the WISE W1 and W2 images (the most affected) by using circular aperture photometry as prescribed by the WISE guide.

The routine of transforming WISE original flux units (digital number, DN) into physical units involves the use of equation \ref{eq:fluxsrc} where $F_{\rm src}$ is the flux of the source (in DN), $f_{\rm apcor}$ is the aperture correction factor respective to each band, $F_{\rm tot}$ is the total flux from the elliptical aperture containing $N_{A}$ number of pixels in the galaxy and $\bar{B}$ is a measure of the background level, also in DN, estimated from the neighbouring region around the source annulus. As a final step to obtain the fluxes, we convert DN units into Jansky~(Jy) using a conversion factor $c$ as in Equation \ref{eq:cfsrc}:
\begin{equation}\label{eq:fluxsrc}
    F_{\rm src} = f_{\rm apcor}(F_{\rm tot})-N_{A}\bar{B})
\end{equation}
\begin{equation}\label{eq:cfsrc}
    F_{\rm src}\; ({\rm Jy}) = cF_{\rm src}
\end{equation}

The WISE explanatory guide also describes the process to derive an error measurement based on an uncertainty map. These images were not available for retrieval, so we decided to use the magnitude-to-standard-deviation ratio for the AllWISE Source Catalog values in order to estimate UGC~2885's errors in magnitude. This uncertainty approximation relies on the uniformity of the WISE data set. Calculating the magnitude of each band based on ${\rm MAG} = {\rm MAGZP} - 2.5\log_{10}(\rm F_{\rm src})$ with ${\rm MAGZP}$ being the magnitude zero-point, also determined for each individual band, and the uncertainties as mentioned, we compare those values with the available magnitudes in Table \ref{tab:magwise}. Both $c$ and ${\rm MAGZP}$ were determined from the in-band relative system response curve  of the photon count of the WISE telescope, derived from the quantum efficiency of each bandpass \citep{jarrett2011spitzer}.

\begin{table}
\caption{WISE magnitudes of UGC~2885}
\label{tab:magwise}
\begin{tabular}{lrr} 
& This work    & AllWISE      \\
\hline \hline
W1 & $12.43 \pm 0.02$ & $12.31 \pm 0.02$\\
W2 &$13.17 \pm 0.03$ & $12.27 \pm 0.03$ \\
W3 & $9.07 \pm 0.04$ & $9.23 \pm 0.04$\\
W4 & $7.52 \pm 0.12$ & $6.87 \pm 0.11$\\
\hline
\end{tabular}
\tablefoot{Magnitudes are given in the Vega system. In this work the foreground star is masked for filters W1 and W2.}
\end{table}

\subsection{CO(1-0): IRAM}

We are interested in investigating H$_{2}$ content of UGC~2885 and specifically the cold, less dense, non-star-forming molecular gas. We are focused on the ground state rotational transition CO($1-0$) \citep{harris2010co, emonts2014co}. UGC~2885 was observed with the Institut de radioastronomie millimétrique (IRAM) 30~m telescope \citep{baars1987iram} in 2021 between July 1--5, August 25--30, and October 14--15. A $5\arcmin \times 1.6\arcmin$ field,  with a position angle -47.5$\deg$, to encompass the entirety of the H$\alpha$ disc in position switching mode. We acquired the CO($1-0$) line with the E0 receiver centred at 109.5~GHz.  The data were reduced and mapped using the GILDAS CLASS\footnote%
{Grenoble Image and Line Data Analysis Software 
Continuum and Line Analysis Single-dish Software, \url{http://www.iram.fr/IRAMFR/GILDAS}}
package. We select the frequency range from 112.29 to 113.59~GHz of the 4~GHz bandwidth, centred around the CO(1-0) line. The bandwidth utilised is about 3000~km~s$^{-1}$. We subtracted a first-order polynomial baseline from each subscan before Hanning smoothing and binning the spectra. The resultant maps have 5\farcs3 pixels, with a 22\farcs49 FWHM beam, and a spectral resolution of 25 km~s$^{-1}$.

To derive the molecular hydrogen mass, we calculate the CO($1-0$) line luminosity from the source flux. We observe that our datacube is significantly affected by noise, background contribution, and edge effects. Thus, we applied a signal masking routine\footnote{\url{https://github.com/radio-astro-tools/tutorials/tree/master/masking_and_moments}} from the \textsc{radio-astro-tools} code \citep{ginsburg2015radio}, which uses  Python packages \textsc{spectral-cube} and \textsc{astropy.modelling} and is suited to deal with radio data (and its specific units). This method consists of measuring the noise level of the cube based on the pixel-by-pixel standard deviation ($\sigma$), as we assume the noise is constant spectrally except for a short range of channels where significant signal exists above the mean $\sigma$. We applied a sigma-clipping  technique to iteratively remove signal from those channels and calculate the average noise level.

\begin{figure*}
    \centering
    \includegraphics[width=0.4\textwidth]{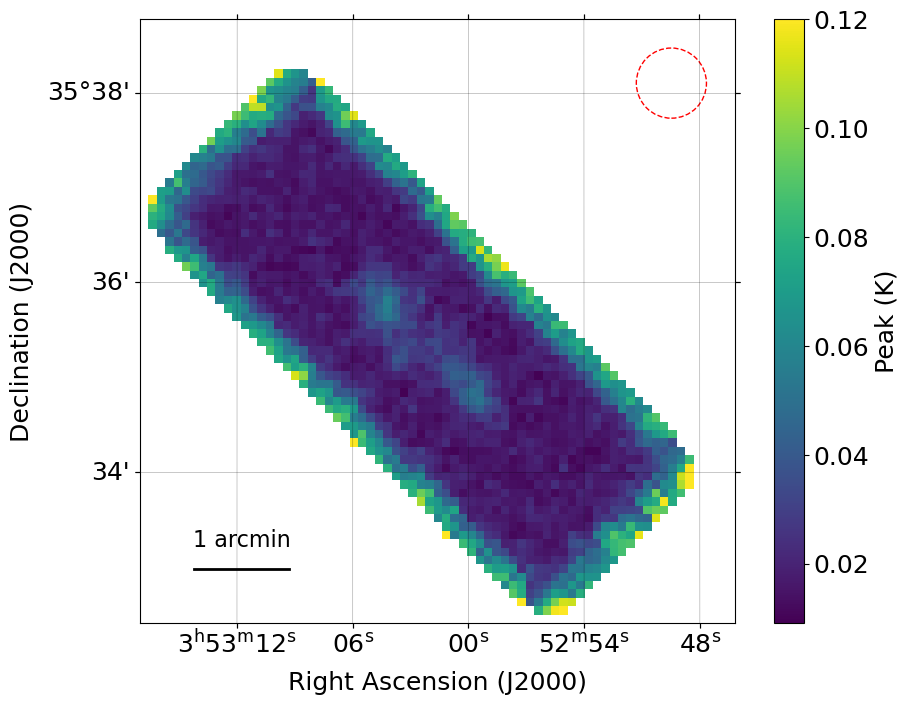}
    \includegraphics[width=0.4\textwidth]{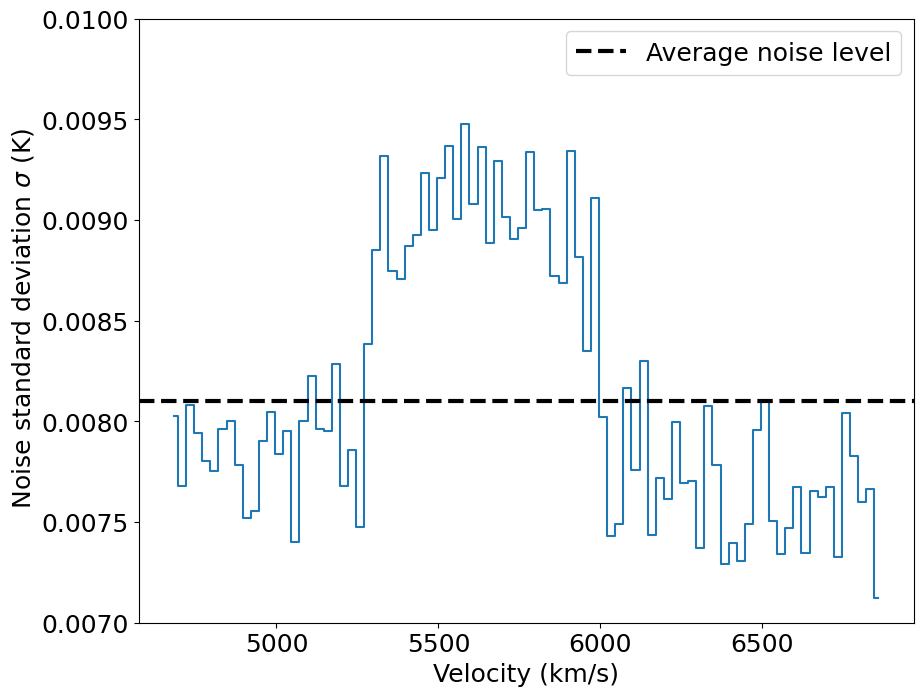}
    \includegraphics[width=0.4\textwidth]{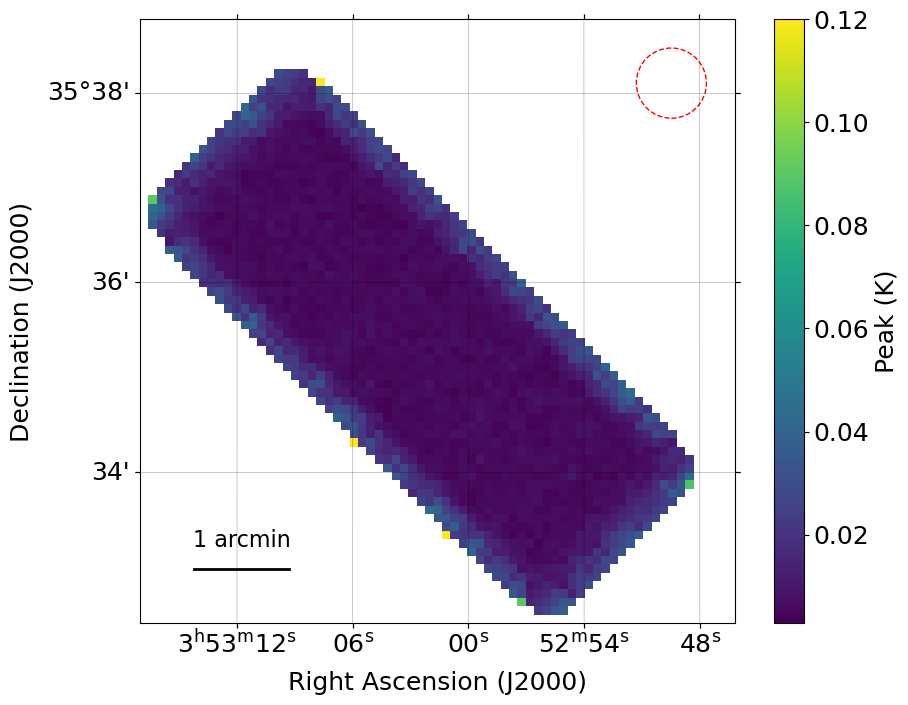}
    \includegraphics[width=0.4\textwidth]{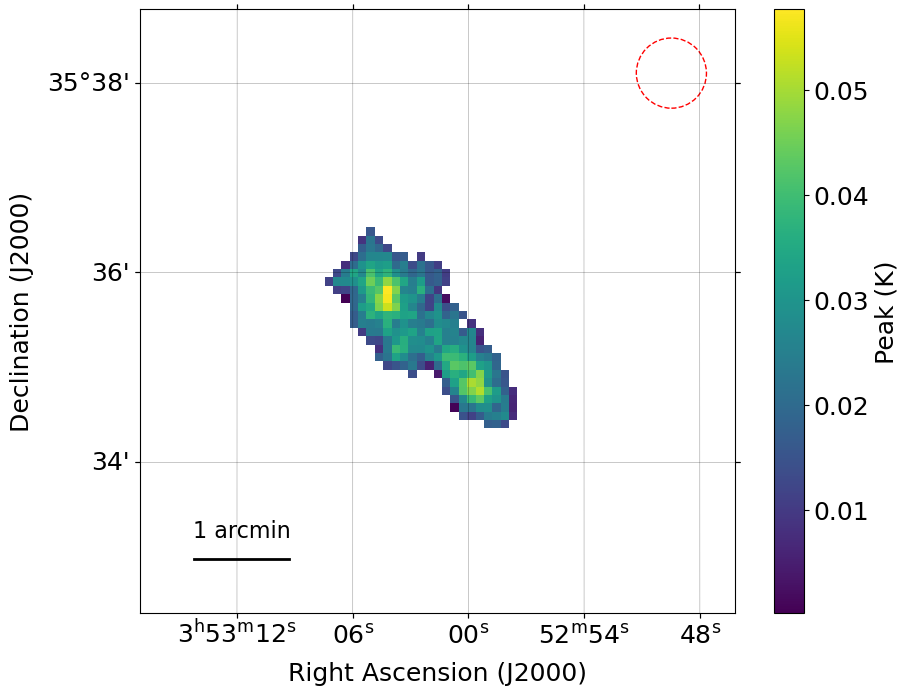}
    \caption{Top-left: Peak intensity of brightness temperature (K) of the original IRAM CO(1-0) cube. Top-right: Noise level $\sigma$ over the spectral range of the cube, black dotted line represents the average level of noise ($\sigma=0.0081\; $~K), used to create the following panel. Noticeably from the central channels, we cannot completely separate the cube's signal from its noise, even after sigma-clipping. Bottom-left: Noise model, or spatial distribution of noise to be subtracted from the original map. Bottom-right: Region with signal that surpasses $5\sigma$. Dashed red circle in the upper right corner shows the $22.49$~arcsec beam size of the IRAM observation.}
    \label{fig:radioastrotools}
\end{figure*}

Here we define two signal masks, low and high masks. The low mask represents the signal that is strongly affected by noise (low SNR) and the high mask stands for the signal that is weakly affected by noise (high SNR). They are defined by the standard deviation level applied on the regions of the map: a region is described as the 26 pixels that are adjacent to a single pixel in the three dimensions. We used $>2\sigma$ and $>5\sigma$ for the low signal and high signal masks, respectively. Another adjustable feature is the number of pixels taken in each region that have to surpass both masks. In our case we decide to adopt a conservative criterion to make sure we are able to identify even relatively noisy features (20 pixels for the low signal mask and 3 pixels for the high signal mask). Figure \ref{fig:radioastrotools} shows the noise model (bottom-left panel) created by the routine as well as the identified signal (bottom-right panel).

\begin{figure}
    \centering
    \includegraphics[scale=0.31]{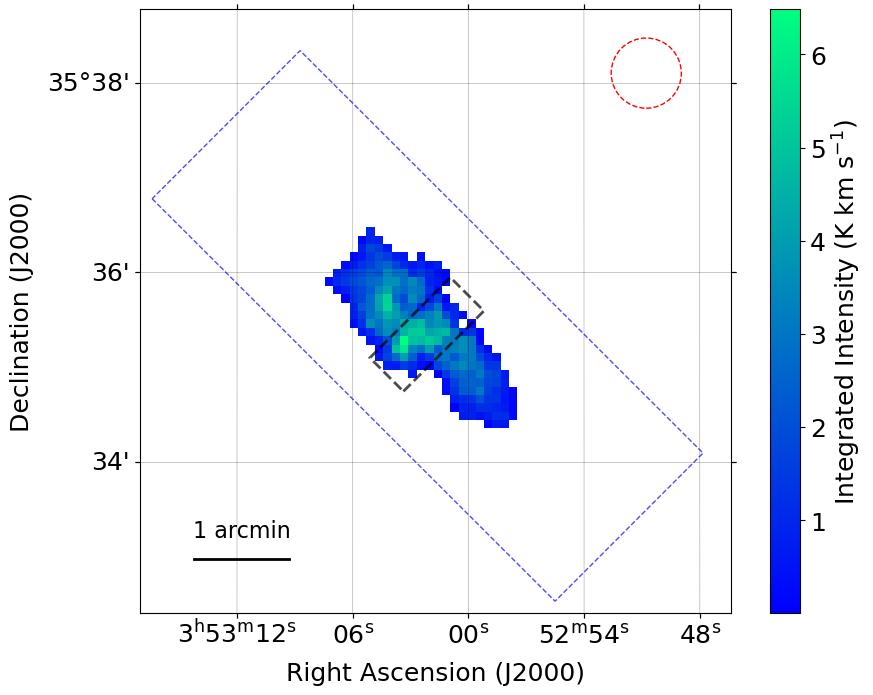}
    \includegraphics[width=0.41\textwidth]{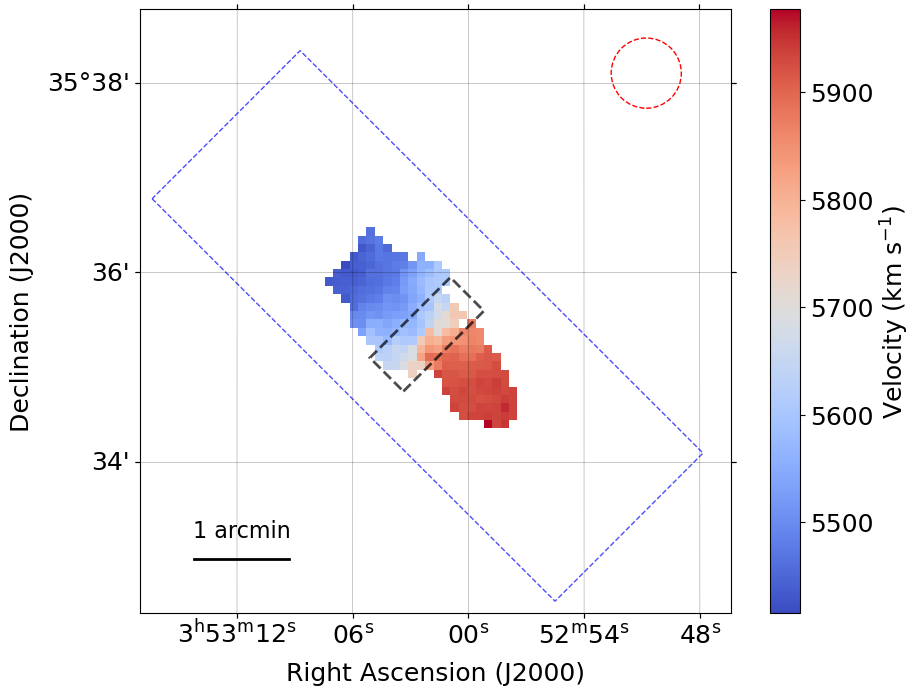}
    \includegraphics[width=0.41\textwidth]{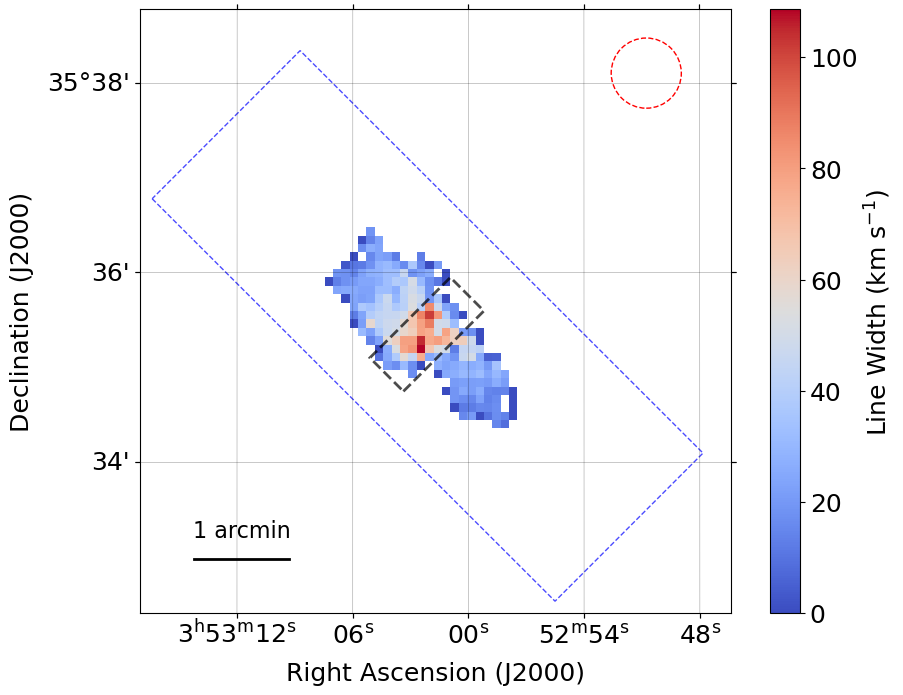}
    \caption{CO moment maps of UGC~2885 from the IRAM datacube. Top: Moment-0 (integrated intensity) map. Middle: Moment-1 (intensity weighted velocity) map. Bottom: Moment-2 (square root of the intensity weighted dispersion, or line width) map. The dashed box in all panels highlights the central region. The box has 1\farcm2 in length, 0\farcm5 in width and it is positioned at the central pixel of the image. The central distribution of molecular gas is discussed further in Section \ref{sec:bar}. Blue box represents the edges of the CO($1-0$) datacube. Red dashed circle is the IRAM beam size.}
    \label{fig:mom1mom2}
\end{figure}

From the output signal map, we derive moment maps that explain the gas dynamics of the system. Moment maps are calculated based on the single Gaussian distribution of the datacube's spectral range.
The zeroth moment (M$_{0}$) is defined as the integrated intensity map, being the integral (or equivalent sum) of the line signal (S($\nu$)) throughout the spectrum where in our case the intensity is in units of K~km~s$^{-1}$. The moment-0 map will be used to calculate the CO($1-0$) line luminosity. 
Moments 1 (M$_{1}$; Equation \ref{eq:moment1}) and 2 (M$_{2}$; Equation \ref{eq:moment2}) highlight different aspects of the gas velocity in the galaxy. M$_{1}$ (shown for completeness but not used in this analysis) measures the centroid velocity of the spectrum peak in each pixel, indicating rotation, while M$_{2}$  traces the velocity dispersion, or more specifically the square of the emission line width. 
\begin{equation}\label{eq:moment1}
    M_{1} = \frac{\int_{line}{vS(v)}dv}{\int_{line}{S(v)}dv} \approx \frac{\sum_{i}v_{i}S(v_{i})\delta v}{M_{0}}
\end{equation}
\begin{equation}\label{eq:moment2}
    M_{2} = \frac{\int_{line}{(v-v_{0})^{2}S(v)}dv}{\int_{line}{S(v)}dv} \approx \frac{\sum_{i}(v_{i}-M_{1})^{2}S(v_{i})\delta v}{M_{0}}
\end{equation}
Moment maps are displayed in Figure \ref{fig:mom1mom2}.

\subsection{Uncertainties for IRAM datacubes}

\begin{figure}
    \centering
    \includegraphics[scale=0.35]{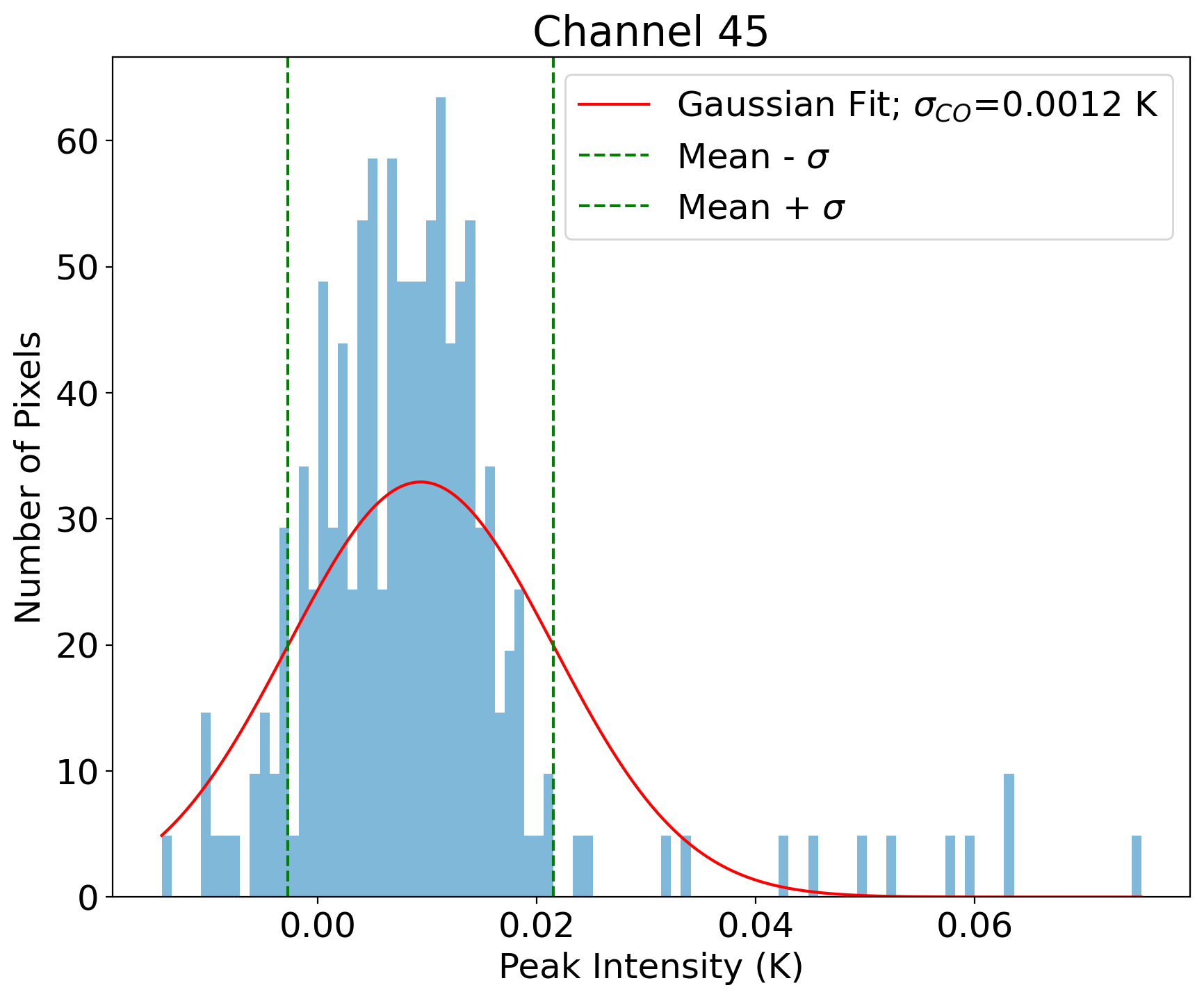}
    \caption{Distribution of brightness temperature in a channel of the datacube. $\sigma_{CO}$ is the channel standard deviation, here taken as the spectral noise. Green dotted lines are spaced $1~\sigma$ from the mean value.}
    \label{fig:histnoise}
\end{figure}

The uncertainties in flux for the radio observations can be estimated based on the line flux emission and line width. We followed the procedure described by \cite{saintonge2017xcold}, with the error in the line flux described in Equation~\ref{eq:errormass} where $\Delta w_{ch}$ is the spectral channel width i.e. channel separation in our datacube ($\Delta \rm w_{ch} = 25$ km~s$^{-1}$), W$_{\rm CO}$ is the line width (estimated pixel-by-pixel from the Moment-2 map of UGC~2885) and $\sigma_{\rm CO}$ the spectral noise. The spectral noise is calculated from the pixel value histogram of a spectral channel of the datacube. We chose a channel with a high signal-to-noise to fit a skewed Gaussian distribution to the negative half of the data. From Figure \ref{fig:histnoise}, we calculate $\sigma_{CO} = 0.0012$~K.

\begin{equation}\label{eq:errormass}
    \epsilon_{obs} = \frac{\sigma_{CO}W_{CO}}{\sqrt{W_{CO}\Delta w_{ch}^{-1}}}
\end{equation}

\subsection{21~cm line: WSRT}

\begin{figure}
    \centering
    \includegraphics[scale=0.36]{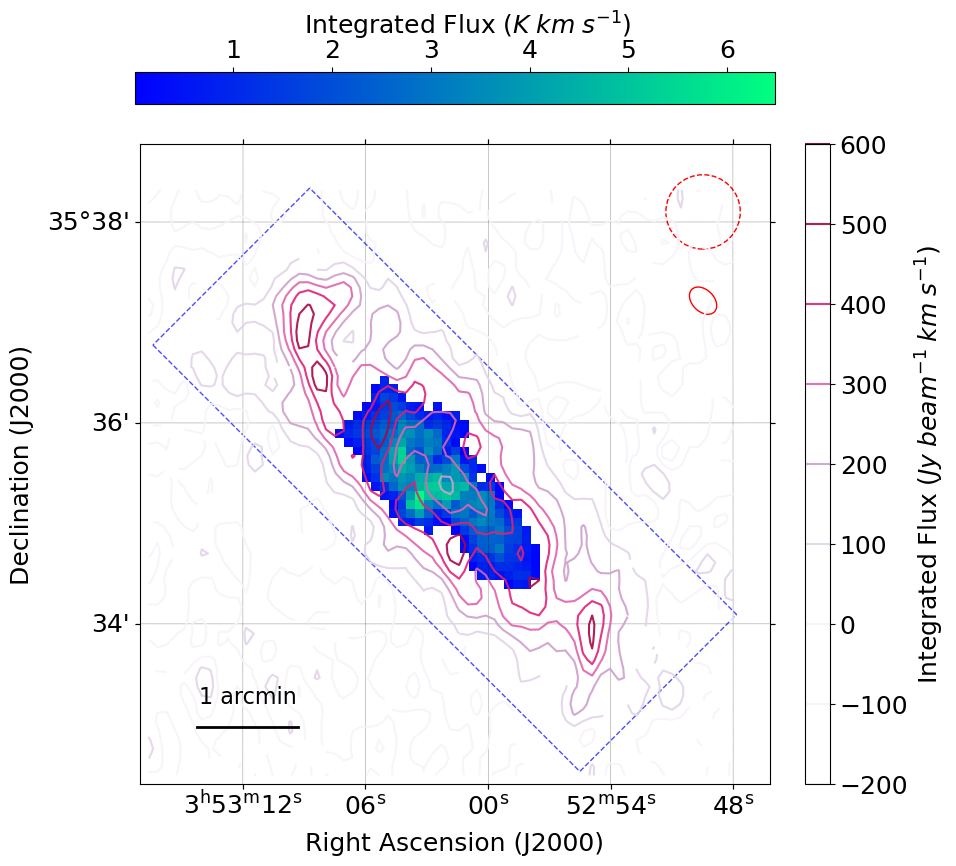}
    \caption{Red-coloured contours represent the distribution of \ion{H}{I} emission in UGC~2885 from the WSRT data. Each increment represents $100$~Jy~ beam$^{-1}~$km$~$s$^{-1}$. The background filled map is the integrated flux (moment-0) from the IRAM datacube. Blue box as described in Figure \ref{fig:mom1mom2}. Red dashed circle is the IRAM beam size and red solid ellipse is the WSRT beam size (13 $\times$ 19~arcsec).}
    \label{fig:contourh1h2}
\end{figure}

In order to understand the evolution of UGC~2885 we also included 21-cm line observations from the Westerbork Synthesis Radio Telescope  taken in 2004, reported by \citet{hunter2013star}, and kindly provided by D. Hunter. For UGC~2885, the images have a spatial resolution of $22\farcs3 \times 13\farcs6$ over a field of $256 \times 256$ pixels, with each pixel  being $4\arcsec$. Figure \ref{fig:contourh1h2} shows the atomic hydrogen distribution compared to the molecular hydrogen in UGC~2885. The blue box was included in this figure to show there is no significant CO($1-0$) emission in the outer parts of UGC~2885, pointing to possible star formation activity driven by atomic hydrogen only.

\section{Estimating the metallicity}
\label{sec:met}

By utilising multiple diagnostics, we aimed to gain a comprehensive understanding of the metallicity of UGC~2885.
A preliminary examination of spectra can be helpful in choosing metallicity diagnostics:
in the UGC~2885 spectra from SITELLE, we observed strong H$\alpha$ and \ion{O}{ii} emission, indicating that UGC~2885 is a high metallicity galaxy, as expected for its mass (and confirmed in Section~\ref{sec:met_results}).
Based on this observation, we have chosen three different metallicity diagnostics, N2O2, R$_{23}$ and O3N2, each offering its own advantages and disadvantages \citep[for a review, see][]{scudder2021}. 

\subsection{N2O2 index}
\label{sec:NIIOII} 

The N2O2 index uses two spectral lines, \ion{N}{II}$\lambda6584$ and \ion{O}{II}$\lambda3727$, which are both unaffected by underlying stellar populations \citep{2002ApJS..142...35K} and expected to be strong even at low signal-to-noise.
This index,
\begin{equation}
\label{eq:n2o2def}
    R = {\rm N2O2} = \log \bigg(\frac{[N II]\lambda6584}{[O II]\lambda3727} \bigg),
\end{equation}
has well-defined emission lines, lacks local maxima and is independent of the ionisation parameter \citep{2002ApJS..142...35K, Paalvast_2017}. It is reported to be reliable for metallicities above half solar \citep[$Z > 0.5 Z_{\odot}$, $\log{\rm [O/H]} + 12 > 8.6$; ][]{Paalvast_2017}. A disadvantage of this index is that the line ratio is sensitive to dust extinction. However, \citep{2002ApJS..142...35K} found that using the Balmer decrement and a classical reddening curve is sufficient to retrieve a reliable metallicity value. We chose the calibration presented by \citet{2002ApJS..142...35K} to calculate the metallicity with the assumption of $12 + \log[{\rm O/H}]>8.6$, 
\begin{equation}
\label{eq:n2o2met}
 12 + \log[{\rm O/H}]  = \log (1.54020 + 1.26602R + 0.167977R^2) + 8.93. 
\end{equation}

\subsection{R$_{23}$ index}
\label{sec:r23} 

The commonly used R$_{23}$ strong line ratio, initially introduced by \citet{1979MNRAS.189...95P}, makes use of four spectral lines, \ion{O}{II}$\lambda3727$, \ion{O}{III}$\lambda5007$, \ion{O}{III}$\lambda4959$ and H$\beta$.
R$_{23}$ is defined as 
\begin{equation}
\label{eq:r23def}
    x = \log {\rm R_{23}} = \log \bigg(\frac{[OII] \lambda 3727 + [OIII] \lambda \lambda 4959, 5007}{ H \beta}\bigg).
\end{equation}
This diagnostic strongly depends on the ionisation parameter and also contains a maximum slightly below the solar value \citep{2002ApJS..142...35K}. 
R$_{23}$ is sensitive to abundance but is also double-valued as a function of metallicity, which causes a problem in determining which solution branch best applies \citep{2002ApJS..142...35K}. We used the N2O2 index results, discussed in more detail in Section~\ref{sec:met_results}, to determine that the high metallicity branch was preferred. We chose the ZKH94 calibration, which is an average of three previous calibrations \citep{1986ApJ...307..431D,1985ApJS...57....1M,1984MNRAS.211..507E} and  works best in the metal-rich regime $12 + \log[{\rm O/H}]  > 8.35$ \citep{2004ApJ...617..240K}:
\begin{equation}
\label{eq:r23met}
12 + \log[{\rm O/H}]  = 9.265 - 0.33x - 0.202x^2 - 0.207x^3 - 0.333x^4. 
\end{equation} 

\subsection{O3N2 index}
\label{sec:O3N2} 
The O3N2 index \citep{1979A&A....78..200A} is a valuable diagnostic for metallicity as its four spectral lines (\ion{O}{III}$\lambda5007$, H$\beta$, H$\alpha$, \ion{N}{II}$\lambda6583$) are easily measurable and exhibit minimum wavelength differences between the line pairs, thus minimising the effect of dust reddening \citep{ho_metallicity_2015,Paalvast_2017}.
O3N2 is defined as
\begin{equation}
\label{eq:o3n2def}
{\rm O3N2} = \log \bigg(\frac{[OIII]\lambda5007}{H\beta} \frac{H\alpha}{[NII]\lambda6583} \bigg).
\end{equation}
\cite{Paalvast_2017} found this diagnostic to be sensitive to metallicity in the range $8.12 < 12 + \log[{\rm O/H}] < 9.05$, which makes it functional in the high metallicity regime. 
As this diagnostic is unaffected by reddening, we chose not to correct the flux maps for reddening. We do not depend on this metallicity evaluation alone because the ionisation energies for \ion{N}{ii} and \ion{O}{iii} are very different; therefore this index is sensitive to changes in energy input from the massive stars \citep{Paalvast_2017}. 
We used the calibration from  \citet{2004MNRAS.348L..59P}:
\begin{equation}
\label{eq:o3n2met}
  12+ \log[{\rm O/H}] = 8.73 - 0.32 ({\rm O3N2}).
\end{equation}
This calibration is useful when $-1 < {\rm O3N2} < 1.9$, and becomes much less reliable beyond ${\rm O3N2} \geq 2$  \citep{2004MNRAS.348L..59P}. 

\subsection{Metallicity uncertainties}

The spatially-resolved metallicity was estimated by combining line flux maps and their corresponding uncertainty maps in the three different indicators as described in the previous subsections. 
Metallicities within each of the ellipses described in Section~\ref{sec:sitelle_data} were computed by averaging the index values over the ellipses.
To calculate the metallicity uncertainties, we created uncertainty maps
by computing the maximum and minimum values of the relevant line ratios, i.e. 
$(a/b)_{\max} = (a+ \delta a)/(b-\delta b)$, $(a/b)_{\min} = (a- \delta a)/(b+\delta b)$, propagating these through the appropriate definitions (Eqs.~\ref{eq:n2o2def}, \ref{eq:r23def}, \ref{eq:o3n2def}) and then using them to compute maximum and minimum metallicities $Z_{\max},Z_{\min}$ at each position (Eqs.~\ref{eq:n2o2met}, \ref{eq:r23met}, \ref{eq:o3n2met}).
The maximum absolute deviation at each position, $\max(Z_{\max}-Z,Z-Z_{\min})$, was taken as 
the metallicity uncertainty.
Averages of uncertainty maps within each ellipse provided the quote uncertainty values.

\section{Estimating star formation rates}

Here we calculate SFR for UGC~2885 based on WISE mid-infrared W3 and W4 luminosities, calculated as described in Equation 3 of \citet{jarrett2012extending}\footnote{L$_{\rm band}$ is the `in-band' luminosity, defined by L$_{\rm band} (\rm L_{\odot}) = 10^{-0.4(\rm M (band) - M_{\odot} (\rm band))}$.}. \citet{cluver2014galaxy, cluver2017calibrating} explored the correlation between mid-infrared WISE bands and star formation rates while studying nearby galaxy surveys such as GAMA \citep{hopkins2013galaxy}, SINGS \citep{kennicutt2003sings} and KINGFISH \citep{kennicutt2011kingfish}. We apply the relations:
\begin{equation}\label{eq:w3}
    \log({\rm SFR})~(M_{\odot}  {\rm yr}^{-1})  = 0.889\pm0.018 \log( L_{\rm W3}) - 7.76\pm0.15
\end{equation}
and
\begin{equation}\label{eq:w4}
    \log({\rm SFR})~(M_{\odot}  {\rm yr}^{-1}) = 0.915\pm0.023 \log(L_{\rm W4} ) - 8.01\pm0.20.
\end{equation}
where WISE luminosities are measured in L$_{\odot}$ and star formation rates in  M$_{\odot}$~yr$^{-1}$. \citet{cluver2017calibrating} discussed the possible effects of metallicity on the luminosity-SFR relations and found that there is little to no correlation between L$_{\rm W3}$ and $Z$. The same effects for L$_{\rm W4}$ were not addressed.
The contaminating effects of the foreground star on W3 and W4 measurements are expected to be smaller than the uncertainties in the SFR calibration.

\section{Estimating gas and stellar masses}
\label{sec:masses}

\subsection{Conversion factor $\alpha_{CO}$}

As a final step to calculate the molecular hydrogen mass, we apply a conversion to the CO($1-0$) line luminosity. The CO-to-H$_{2}$ conversion can be found in \citet{sandstrom2013co} as the relation of Equation \ref{eq:conversion} with $\alpha_{\rm CO}$\footnote{also denoted $X_{\rm CO}$, in units of  cm$^{-2}$~(K~km~s$^{-1}$)$^{-1}$.} in units of M$_{\odot}$(K~km~s$^{-1}$~pc$^{2}$)$^{-1}$ and being determined based on the column density of the molecular gas, L$_{\rm CO}$ is the integrated luminosity of the CO(1-0) line in units of L$_{\odot}$. In this study we adopt $\alpha_{\rm CO}=4.3$, based on the observations of \citet{strong1996gradient} and the application of the model to giant molecular clouds in the Milky Way, explored in detail by \citet{bolatto2013co}. Here we assume the Milky Way's metallicity and a radially constant value for $\alpha_{\rm CO}$ over the galaxy's extent.
\begin{equation}\label{eq:conversion}
    M_{H_{2}} \;(M_{\odot}) = \alpha_{CO} L_{CO}
\end{equation}
Applying this factor has some implications for UGC~2885's molecular properties. The conversion between CO emission and molecular hydrogen mass is thought to depend on metallicity and to also have a radial dependence \citep{arimoto1996co}. As metallicity increases, $\alpha_{\rm CO}$ decreases, since by definition we see a lower hydrogen abundance in relation to oxygen \citep{hirashita2023effects, sandstrom2024resolved}. We discuss the effects of the estimated metallicities of this study in calculating M$_{\rm H_{2}}$ for UGC~2885 in Section~\ref{sec:m2}.

\subsection{Neutral hydrogen mass}
The \ion{H}{I} mass is calculated based on the Equation \ref{eq:conversionhi} where F$_{\rm HI}$ is the integrated flux of the \ion{H}{I} line, in units of Jy~km~s$^{-1}$. We estimate the uncertainties for the \ion{H}{I} mass based on the relations described by \cite{doyle2006effect}:

\begin{equation}\label{eq:conversionhi}
    M_{HI} \;(M_{\odot}) = 2.356 \times 10^{5} D_{L}^{2} F_{HI} 
\end{equation}
\begin{equation}
    \Delta F_{HI} = 0.5 F_{HI}^{1/2}
\end{equation}
\begin{equation}
    \Delta M_{HI} = M_{HI}\frac{\Delta F_{HI}}{F_{HI}}.
\end{equation}

\subsection{Stellar mass-to-light ratio}

To compute the stellar mass (M$_{*}$), we follow the procedure described by \cite{parkash2018relationships}. This relation found in Equation \ref{eq:conversionmstar} uses the W1 in-band luminosity with M$_{\rm Sun} = 3.24 $ \citep{jarrett2012extending} and M as the absolute magnitude of the WISE 1 band:
\begin{equation}\label{eq:conversionmstar}
    L_{W1} \; (L_{\odot}) = 10^{-0.4(M-M_{Sun})}.
\end{equation}
The mass-to-light ratio ($\Upsilon_{\star} = \rm M_{*}/L$) of a galaxy depends on various properties of the object, such as its morphology and colour. Multiple studies have explored the different equivalences between emitted light and stellar content using WISE colours \citep[most notably W1--W2 and W1--W3; ][]{jarrett2012extending, meidt2014reconstructing, kettlety2018galaxy, parkash2018relationships}). Here we adopt $\rm M_{*}/L_{W1} = 0.35 \pm 0.11 \; \rm  M_{\odot}/L_{\odot}$ as \citet{jarrett2023new} find a tight linear relation between stellar mass and W1 luminosity.

\section{Results and discussion}\label{sec:results}

\subsection{Metallicity and metallicity gradient}
\label{sec:met_results}

\begin{figure}
\centering
	\includegraphics[scale=0.31]{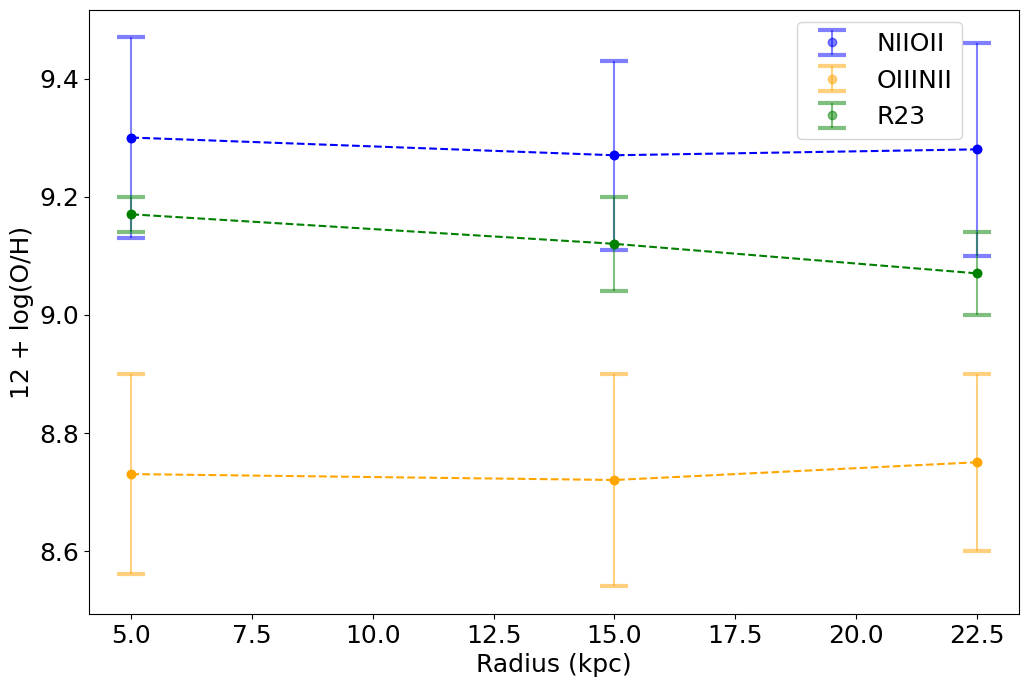}
    \caption{Metallicity  calculated using N2O2, R23 and O3N2 diagnostics.
    The specific radial zones evaluated were  0--10~kpc (ellipse A), 10--20~kpc (ellipse B), and 20--25~kpc (ellipse C).
    The overall metallicity value agrees well between N2O2 and R23, with the O3N2 metallicity offset to a lower value. 
    None of the three indicators shows a strong metallicity gradient.}
    \label{fig:metallicityplotvalue}
\end{figure}

Using N2O2, R23 and O3N2, the global metallicity was calculated within a radius of 25 kpc to be $Z = 9.28, 9.08$ and 8.74, with uncertainties of $0.17$ dex, $0.07$ dex, and $0.16$ dex, respectively. 
All three global metallicities show that UGC~2885 is in the high metallicity regime, consistent with our expectations based on the mass-metallicity relationship.
The N2O2 and R23 measurements agree within their uncertainties, while the O3N2 value is slightly lower; possible explanations include the greater sensitivity of O3N2 to variations in the input ionizing spectrum, or the true metallicity of the galaxy being above the validity range of this indicator (see Section~\ref{sec:O3N2}).
We further assessed the metallicity across three distinct zones within UGC~2885's disc to determine its metallicity gradient, as illustrated in Figure \ref{fig:metallicityplotvalue}. 
(As the measured metallicities are well above the lower limit for which the calibrations are valid,
we do not expect our original assumption of high metallicity to affect measurements of the metallicity gradient.)
Within the uncertainties, none of the three indicators shows a strong metallicity gradient.
Table \ref{tab:metallicitychart} gives the metallicities and gradients determined with N2O2, R23 and O3N2 diagnostics. 

Local galaxies often exhibit negative metallicity gradients, with typical values  from 0 to $-0.1$~dex~kpc$^{-1}$ \citep{Sharda_2021}.
The negative pattern arises as star formation in galaxies often starts at the core and expands outwards. Consequently, the central region tends to be more metal-rich compared to its outskirts. Isolated galaxies that do not undergo mergers tend to evolve smoothly, allowing steady star formation resulting in enrichment, but also creating a shallower metallicity gradient \citep{Tissera_2021}. 
Our prediction for an isolated galaxy of this stellar mass would be a shallow negative or flat metallicity gradient accompanied by high average metallicity \citep{Mannucci_2010}, and
our observations are consistent with this prediction.

\begin{table}
        \caption{The metallicity gradient observed in UGC~2885}
    \begin{tabular}{l c c c r} 
    \hline
         \hline
         Diag. & \makecell{$Z_{A} \pm \delta  Z_{A}$ \\(dex)} & \makecell{$Z_{B} \pm \delta  Z_{B}$\\(dex)} &\makecell{ $Z_{C} \pm \delta  Z_{C}$ \\(dex)} &\makecell{ gradient \\dex~kpc$^{-1}$}\\  
        \hline
        N2O2 & $9.30 \pm 0.17$  & $9.27 \pm 0.16$  & $9.28 \pm 0.18$ & $-0.0004$\\  
         
        R23 & $9.17 \pm 0.03$  & $9.12 \pm 0.08$  & $9.07 \pm 0.17$ & $-0.002$  \\
         
        O3N2 & $8.73 \pm 0.17$  & $8.72  \pm 0.18$  & $8.75 \pm 0.16$& 0.0004\\
    
    \end{tabular}
\tablefoot{Ellipse A covers 0--10~kpc, B covers 10--20 kpc, and C 20--25 kpc.}
    \label{tab:metallicitychart}
\end{table}

\subsection{Star formation rate}\label{sec:sfr}

The SFR was estimated, from equations \ref{eq:w3} and \ref{eq:w4}, respectively as $1.41 \pm 0.72$~M~$_{\odot}$~yr$^{-1}$ and $1.86 \pm 1.24\; \rm{M}_{\odot} \; \rm{yr}^{-1}$, giving a mean value of $1.63 \pm 0.72$~M~$_{\odot}$~yr$^{-1}$. This agrees with the calculated $2.47$~M~$_{\odot}$~yr$^{-1}$ of \cite{hunter2013star} within the uncertainties of this work and assuming a similar uncertainty value attributed to their measurement. However, \citet{hunter2013star} used a different estimated distance for UGC~2885, D$_{\rm L} = 79.10$~Mpc. If those authors had used D$_{\rm L} = 84.34$~Mpc their SFR would increase to 2.80~M$_{\odot}$~yr$^{-1}$. The divergence could be attributed to the star formation indicators used: H$\alpha$ emission probes star formation on shorter timescales than UV or infrared emission \citep{ke2012}.

Considering UGC~2885's size, we are also interested in calculating its SFR surface density ($\Sigma_{\rm {SFR}}$). In this work the surface densities are estimated at the outer radius of the largest ellipse in Fig.~\ref{fig:fluxmap}, 25~kpc. $\Sigma_{\rm {SFR}}$ is calculated as 1.97$\times$10$^{-3}$~M$_{\odot}$~yr$^{-1}$~kpc$^{-2}$ and is comparable to other giant galaxies in the nearby Universe, ranging from 0.50-2.16$\times$10$^{-3}$~M$_{\odot}$~yr$^{-1}$~kpc$^{-2}$ \citep{ray2024probing}. This result is surprising considering that in \citet{ray2024probing} they have spatially resolved measurements of SFR, which is the main caveat of this comparison.

\subsection{Molecular hydrogen mass and star formation efficiency}\label{sec:m2}

We find a molecular hydrogen mass of $1.89 \pm 0.24 \times 10^{11} \; \rm M_{\odot}$. This represents the first estimate of the H$_{2}$ content of UGC~2885 and will allow us to calculate the Star Formation Efficiency (SFE) of this galaxy. SFE can be derived from the relation SFE~(yr$^{-1}$) = ${{\rm SFR}/{\rm M_{H_{2}}}}$ and has a mean value of $8.67 \pm 4.20 \times 10^{-12}$~yr$^{-1}$ over the mean calculated star formation rate from Section~\ref{sec:sfr}. Nearby disc galaxies of high mass (M$_{\star} >10^{10}~\rm M_{\odot}$) have a total molecular gas mass that ranges from $\sim10^{8}$ to $\sim10^{10}~\rm M_{\odot}$, indicating that UGC~2885 is an unique object with a large gas reservoir \citep{saintonge2017xcold}. Moreover, nearby discs have mean SFE of 5.25$\pm 2.50\times 10^{-10} \; \rm{yr}^{-1}$ \citep{leroy2008star}, which makes UGC~2885 extremely inefficient in forming stars from its available molecular gas. We estimate an upper limit on the integrated intensity of the CO(1-0) line where we could not find a detection up to 50~kpc in radius. Signal over 3$\sigma$ = 0.12~K~km~s$^{-1}$~beam$^{-1}$ would be evidence of a detection.

We also study the effect of correcting the conversion factor $\alpha_{\rm CO}$ for metallicity on the estimation of M$_{\rm H_{2}}$. \citet{bolatto2013co} presented a way to calculate the $\alpha_{\rm CO}$ that is dependent on both the galaxy's metallicity and the total surface density. We see this relation in Equation \ref{eq:convfmet}\footnote{This is a modified version that avoids incongruous $\alpha_{\rm CO}$ values in low surface-density regions \citep{sun2023star}.}. Here we use $Z'$ as the metallicity scaled by the solar metallicity \citep[$Z'$ = 0.014 or $12 + \log (\rm{{O}/{H}} = 8.7)$; ][] {asplund2009chemical}. $\Sigma_{\star}$ and $\Sigma_{\rm mol}$ are the surface densities of the stellar and molecular hydrogen masses, respectively. If the sum of these two values is $\geq 100$~M$_{\odot}$~pc$^{-2}$, then $\gamma = 0.5$, otherwise $\gamma = 0$.

\begin{equation}\label{eq:convfmet}
    \alpha_{CO} = 2.9 * \exp \left(\frac{0.4}{Z'}\right) \left(\frac{\Sigma_{\star} + \Sigma_{mol}}{100~M_{\odot} \; {\rm pc}^{-2}}\right)^{-\gamma}
\end{equation}
$\Sigma_{\star} + \Sigma_{\rm mol}$ is calculated as 8.10$\times10^5$~M$_{\odot}$~pc$^{-2}$ --- details on the stellar mass calculation are discussed below in Section \ref{sec:stellarmass}. We estimate $\alpha_{\rm CO}$ as 2.91 for all metallicity indexes presented in this work. Using this conversion factor we calculate the molecular gas mass as M$_{\rm H_{2}} = 1.26 \pm 0.16 \times 10^{11}$~M$_{\odot}$. In this work, we utilise the uncorrected molecular gas mass value, as our primary focus is on comparing this measurement to other studies, some of which do not implement the same corrections. The primary study we compare with in Section \ref{sec:mainseq} applies a correction based on a property we directly investigate.

\subsection{Neutral hydrogen mass}

The atomic hydrogen mass of UGC~2885 is calculated to be M$_{\rm HI} = 3.73 \pm 0.38 \times 10^{10}$~M$_{\odot}$. This value can be compared to the one found in \cite{hunter2013star}, M$_{\rm HI} = 4.20 \pm 0.42 \times 10^{10} \; \rm{M}_{\odot}$. If this mass was measured using D$_{L} = 84.34$~Mpc, the previous authors would have obtained M$_{\rm HI} = 4.33 \pm 0.43 \times 10^{10} \;\rm{M}_{\odot}$. M$_{\rm HI}$ remains consistent between our calculations and the available estimate. 

\subsection{Stellar mass}\label{sec:stellarmass}

We estimate UGC~2885's stellar mass as M$_{\star} = \rm L_{\rm W1}*\Upsilon_{\star} = 4.83 \pm 1.52 \times 10^{11} \; \rm M_{\odot}$. \citet{di2023dark} calculated the same property as $1.58 \pm 0.73 \times 10^{11} \; \rm  M_{\odot}$ using D$_{\rm L} = 71$~Mpc. M$_{\star}$ does not agree within the uncertainties with the most recent published value. Although \cite{di2023dark} used the same available WISE images to estimate stellar mass, they chose a larger mass-to-light ratio than ours ($\Upsilon_{\star} = 0.6$), which could be the reason for this divergence. If those authors had applied our value of D$_{\rm L} = 84.34$~Mpc and a $\Upsilon_{\star} = 0.35 \pm 0.11$~M$_{\odot}$/L$_{\odot}$, they would have estimated M$_{\star} = 1.58\pm 0.84 \times 10^{11} \; \rm M_{\odot}$, which is even more discrepant. The fixed mass-to-light ratio used by \citet{di2023dark} might be more appropriate for elliptical or spheroidal galaxies with little star formation \citep{meidt2014reconstructing}. The $\Upsilon_{\star}$ used in this work also takes into account the WISE colours, reflective of star forming processes such as warm dust in the interstellar medium \citep{leroy2019z}.

\subsection{UGC~2885's position on the star forming main sequence of galaxies}\label{sec:mainseq}

We explore the relationship between SFR, H$_{2}$ mass, \ion{H}{I} mass and stellar mass for UGC~2885. Studies with simulated or observed galaxies have showed the strong correlation between stellar mass and SFR in nearby galaxies, the so-called star forming main sequence \citep[SFMS; ][] {brinchmann2004physical}.

\begin{figure*}
    \centering
    \includegraphics[width=0.45\textwidth]{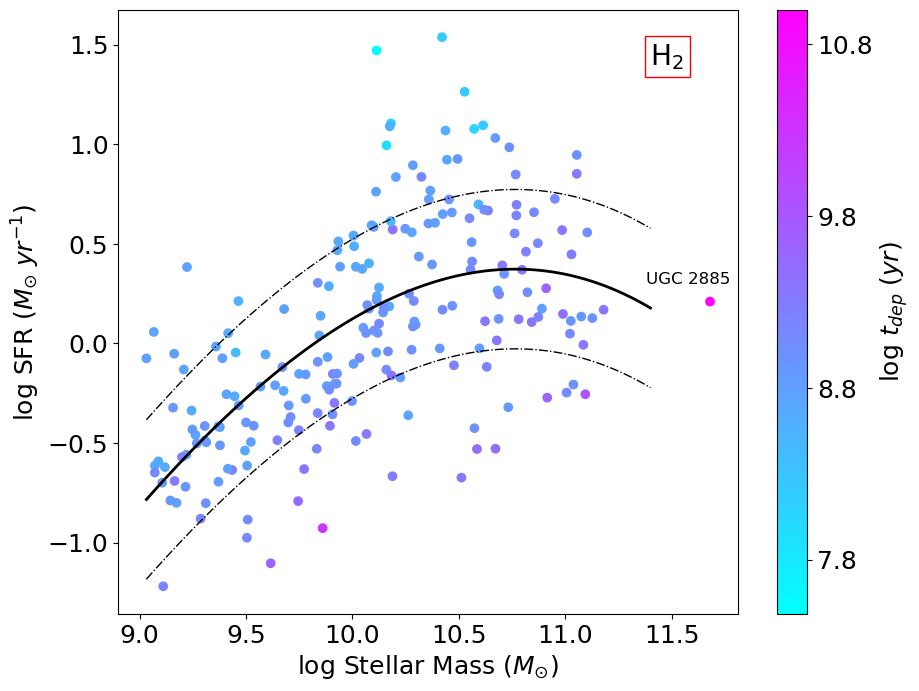}
    \includegraphics[width=0.45\textwidth]{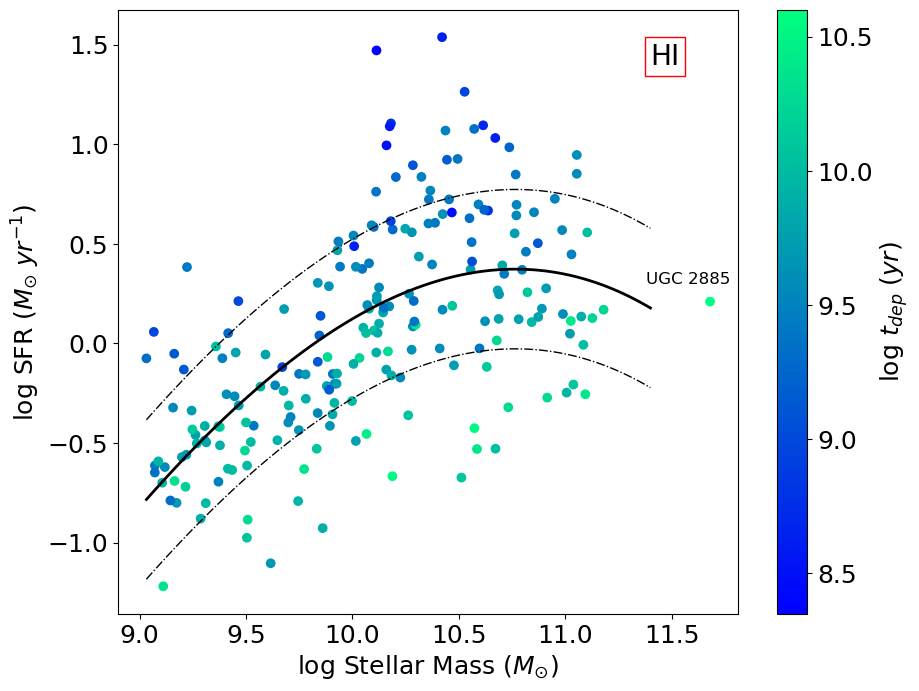}
    \caption{Star forming main sequence (SFMS) of the xCOLD GASS sample with available ALFAFA \ion{H}{I} fluxes. Filled line is the SFMS and dashed lines show $\pm $0.4~dex scatter as described by \cite{saintonge2017xcold}. Colour represents the depletion time (t$_{\rm dep}$) of the molecular gas (left panel) and atomic gas (right panel).}
    \label{fig:depms}
\end{figure*}

The xCOLD GASS survey explored the SFR-M$_{\star}$ plane for local galaxies with available IRAM 30m observations \citep{saintonge2017xcold}. Our comparison to their sample would place UGC~2885 along the distribution but at the higher end of stellar mass \citep[see Figures 7 and 8 of ][]{saintonge2017xcold}. This prediction is reasonable as we expect a flattening of the SFR with a decrease of the available gas reservoir. They also discuss the depletion time for the gas of their sample. Gas depletion time is defined as the time it takes for the gas to be consumed (by either being transformed into stars or accreted by the central black hole) and it can be estimated from the SFE as t$_{\rm dep} = 1/{\rm SFE}$ for both molecular and atomic hydrogen gas. For UGC~2885, the depletion times are t$_{\rm dep}~\rm{(H_{2})} = 1.15 \pm 0.51 \times 10^{11}$~yr and t$_{\rm dep}~\rm{(HI)} = 2.29 \pm 1.04 \times 10^{10}$~yr. Figure \ref{fig:depms} shows the SFR - M$_{\star}$ plane for the xCOLD GASS sample colour-coded by the depletion times, as well as the position of UGC~2885 on it.

\begin{figure}
    \centering
    \includegraphics[width=0.45\textwidth]{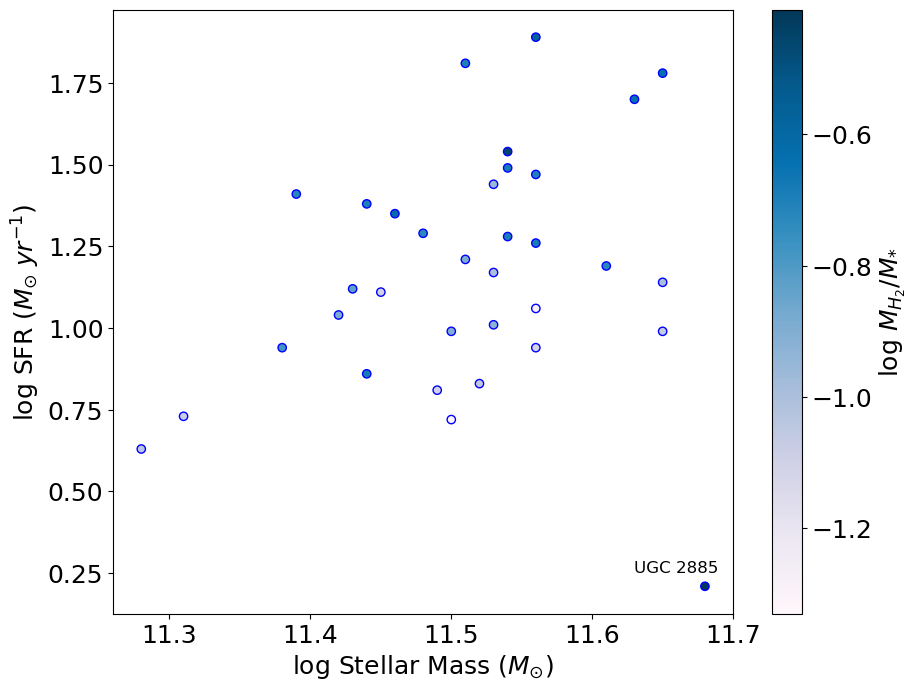}
    \caption{SFR as a function of stellar mass for a sample of massive disc galaxies \citep{lisenfeld2023molecular} colour coded by their molecular-to-stellar mass ratio ($\rm{M}_{H_{2}}/\rm{M}_{\star}$). For comparison, UGC~2885 is displayed but not included in the fitting of the SFR-M$_{\star}$ plane. The range of stellar masses differs from Figure \ref{fig:depms}, as super spiral galaxies have $\log M_{\star} > 10.47$~M$_{\odot}$.}
    \label{fig:lisenfeldsfms}
\end{figure}

When comparing to super spiral galaxies, UGC~2885 sits far above the representative curve due to its high molecular-to-stellar mass ratio ($\log \;f_{\rm H_{2}} = \rm M_{\rm H_{2}}/ \rm M_{\star} = -0.41 \pm 0.33$). $f_{\rm H_{2}}$ for UGC~2885 deviates from higher mass objects (>log~M$_{\star} = 10.47$~M$_{\odot}$), as their molecular-to-stellar mass ratio tends to be much lower \citep[mean value of $\log \;f_{\rm H_{2}} = -1.36 \pm 0.02$; ][] {lisenfeld2023molecular}. Figure \ref{fig:lisenfeldsfms} shows the representative curve of the SFR-M$_{\star}$ plane for super spirals compared to UGC~2885.

\subsection{Fundamental metallicity relation}

\begin{figure*}
\centering
\sidecaption
    \includegraphics[width=12cm]{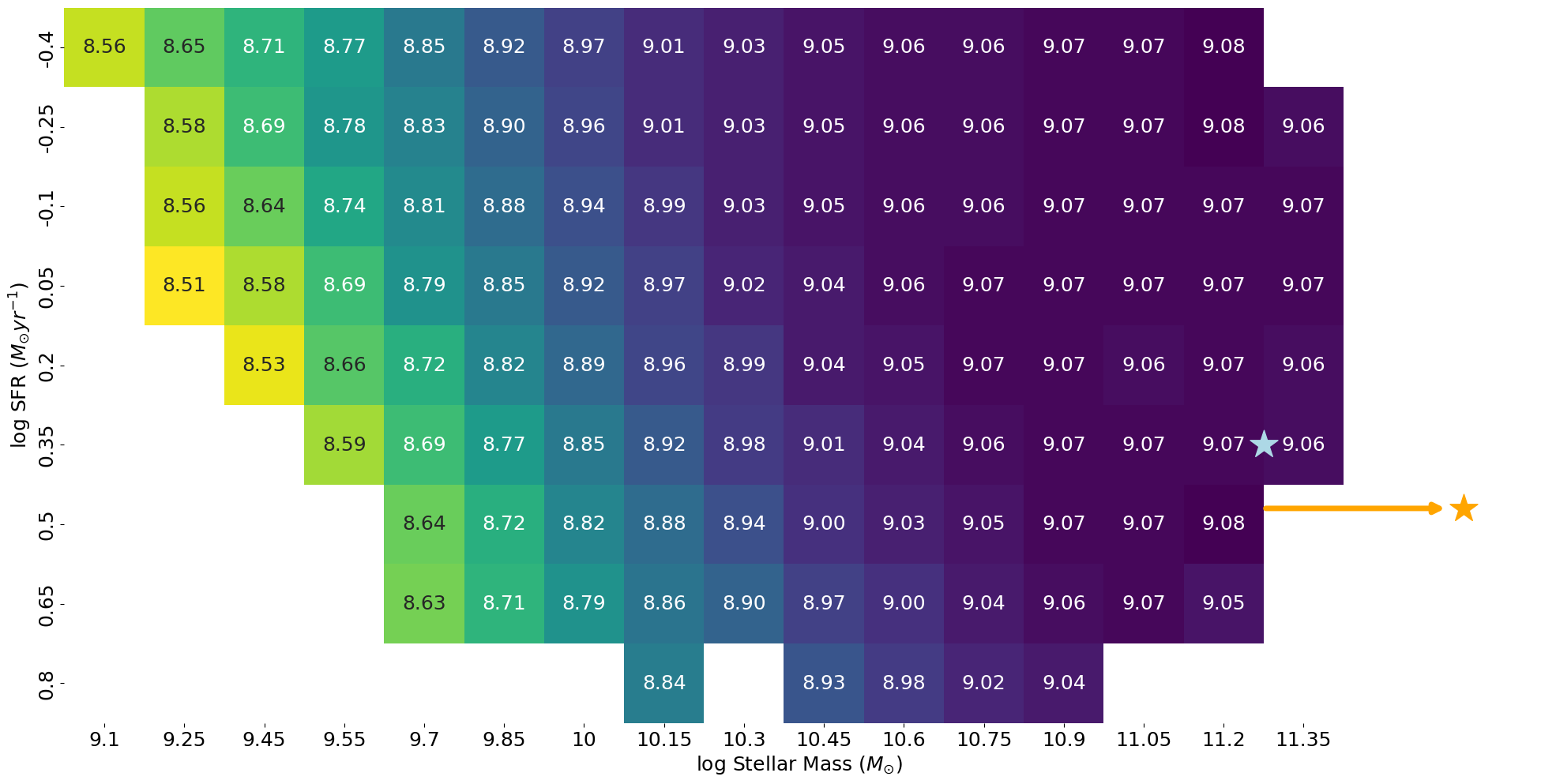}
    \caption{Relation between galaxy stellar mass, star formation rate and metallicity based on data given by \cite{Mannucci_2010}. The orange star represents  UGC~2885's position in the chart based on our calculated values: stellar mass $\log(\rm M_{\star}) = 11.68$ and star formation rate $\log(\rm SFR) = 0.48$. The blue star represents UGC~2885's position in the chart based on previous values \citep[$\log(\rm M_{\star}) = 11.3$, $\log(\rm SFR) = 0.40$;][]{hunter2013star}. Both metallicities found using the FMR are comparable to the results found using the emission line diagnostics to calculate metallicity ($Z \approx 9)$.}
    \label{fig:FMRimage}
\end{figure*}

 The fundamental metallicity relation (FMR) relates stellar mass, gas-phase metallicity ($Z$) and SFR \citep{Mannucci_2010}. Despite its immense size, metallicity and low star formation rate for a massive galaxy, UGC~2885 remains consistent with the FMR. Figure \ref{fig:FMRimage} is the  fundamental metallicity relation recreated using data provided in Table 1 of \cite{Mannucci_2010}, where the values within the bins are the median metallicity found at each given SFR and ${\rm M}_{\star}$. 
 Using the previously published stellar mass and star formation for UGC~2885 \citep[$\log {\rm M}_{\star} = 11.2 \; {\rm M}_{\odot}$, $\log {\rm SFR} = 0.4 \; {\rm M}_{\odot} \; {\rm yr}^{-1}$; ][]{di2023dark,hunter2013star}, the FMR presented in Figure \ref{fig:FMRimage} predicts a metallicity of $Z \approx 9.06$.  
The arrow in Figure \ref{fig:FMRimage} indicates our larger estimated stellar mass and SFR values($\log \rm M_{\star} = 11.68 \; \rm M_{\odot}$, $\log \rm SFR = 0.48 \; \rm M_{\odot} \; yr^{-1}$)
from which the FMR predicts a metallicity of $Z \approx 9.08$.
Both predicted metallicity values are consistent with the global values calculated from our data; within the uncertainties of our measurements, we find that UGC~2885 is not an outlier to the fundamental metallicity relation.

\setlength{\parskip}{3pt}

\citet{Mannucci_2010} found that at large stellar masses ($\log~\rm M_{\star} > 10.9$), there is a lack of correlation between stellar mass and metallicity. At low star formation rates, the correlation also weakens. This is due to the fact that there is a turnover mass ($10^{10} \rm M_{\odot}$) where metallicity begins to saturate. As the metallicity saturates, the mass-metallicity relation also begins to flatten \citep{Zahid_2014}.

\subsection{On the evolution of UGC~2885}

The distance between a galaxy and the star forming main sequence ($\Delta_{\rm MS}$) is an important proxy for galaxy evolutionary stage at the time of observations. Quiescent galaxies, i.e. galaxies with dormant star formation, are generally found to be distant from the MS, specifically below the trend. These galaxies will most likely continue to form stars at a constant but low rate, meaning their global metallicity ceases to change significantly \citep{looser2024stellar}. 

We expect galaxies to increase their metallicity after cycles of star formation and stellar mass growth. Therefore, UGC~2885's high metallicity indicates that the galaxy has cycled through many phases of star formation. This could explain UGC~2885's location in relation to the SFMS, as we see that $\Delta_{\rm MS}$ is inversely proportional to $Z$ \citep{peng2015strangulation} considering that the galaxy will use its gas reservoir to increase its stellar mass, consequently increasing metallicity. The molecular gas fraction $f_{\rm H_{2}}$ is also a main factor to determine $\Delta_{\rm MS}$ since a galaxy will only be star forming with an available gas reservoir. Here we observe a strong discrepancy with the starvation hypothesis as high-metallicity, non star forming galaxies are found to have lower gas fractions than those of MS galaxies. From our measurements, UGC~2885 has a $f_{\rm H_{2}}$ that is comparable to galaxies above the main sequence \citep{saintonge2017xcold}. 
Its molecular gas mass is about three times the maximum M$_{\rm H_{2}}$ of local post-starburst galaxies with comparable star formation rates \citep[see Fig. 3 of][]{french2021},
suggesting that UGC~2885 has not used up its gas reservoir in a recent starburst.

\begin{figure}
    \centering
    \includegraphics[width=0.5\textwidth]{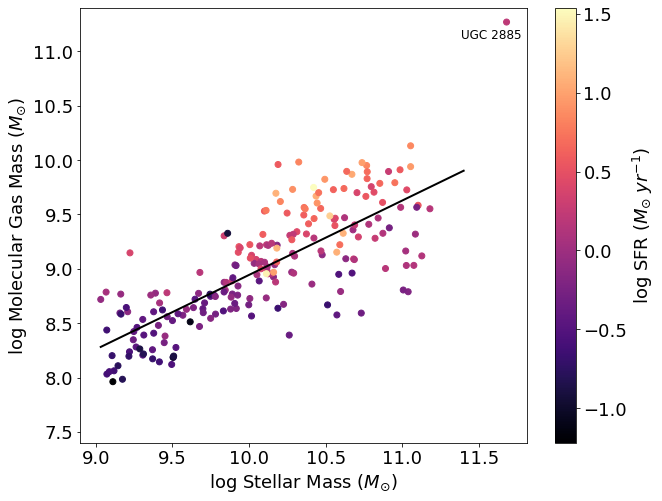}
    \caption{Molecular hydrogen mass as a function of stellar mass (molecular gas main sequence) for the xCOLD GASS sample. The sample is colour-coded by SFR. UGC~2885 is far above the relation, due to its large gas reservoir.}
    \label{fig:mgms}
\end{figure}

The issue with using the SFMS as a probe for galaxy evolution, lies, therefore, on the absence of the galaxy's gas fraction on it. More fundamental relations can be used instead, such as the molecular gas main sequence \citep{lin2019almaquest}. Based on this relation, molecular hydrogen content and stellar mass will follow a linear trend for star forming galaxies. The strong linearity of the M$_{\rm H_{2}}$ - $\rm{M}_{\star}$ plane comes from the gravitational potential cause by either the dark matter content dominance in the galaxy or the baryonic components dominating instead. In both cases, the gas fraction will not affect the trend for low redshifts \citep{baker2023molecular,lin2019almaquest}. Figure \ref{fig:mgms} shows that this might not be the case for UGC~2885 as its current star formation is insufficient to place the galaxy along the linear relation when comparing to galaxies with a high molecular gas reservoir.

\subsection{Could a galactic bar quench UGC~2885's star formation?}\label{sec:bar}

\cite{saintonge2016molecular} discussed the SFMS at the higher end of stellar masses. Specifically, galaxies with M$_{*} > 10^{10.8}\;\rm{M}_{\odot}$ might be part of a `danger zone' (at risk of becoming red and dead). Their vast molecular gas reservoir is not being transformed into stars by some sort of quenching mechanism within the last 1~Gyr. These objects have a high molecular-to-atomic gas ratio, and it is thought that these galaxies have a mechanism that moves gas towards the central region. \citet{saintonge2016molecular} also observes that the majority of the galaxies in the danger zone have strong galactic bars which are potentially strong drivers of gas across the host and towards its centre. We notice that the distribution of molecular gas in UGC~2885 does not expand isotropically from the galactic centre. This is expected for galaxies with galactic bars as seen in the EMPIRE survey (e.g. NGC 2903, NGC 3627 and NGC 6946, \citet{jimenez2019empire}). High-resolution measurements of the molecular hydrogen of UGC~2885 could shed light on its radial distribution, helping to confirm this hypothesis.

It is known that turbulent velocity fields are drivers of shocks in the interstellar medium \citep{hopkins2012structure}. This mechanism can constrain star formation activity on small scales due to its random motion nature, which counteracts self-gravity. Galactic bars induce gravitational torques due to the bar potential--- perpendicular to the structure--- which, then, causes a decrease in the gas rotational motion, meaning an increase in local turbulence. Consequently, we can trace the randomness of the gas by observing the molecular gas dispersion \citep{kim2007gravitational}. 

Figure \ref{fig:mom1mom2} (bottom panel) shows the velocity dispersion of the molecular gas for UGC~2885. A box around the central region of the galaxy highlights where the higher dispersion correlates with the presence of H$_{2}$ gas (see top panel, Figure \ref{fig:mom1mom2}). 
We consider this tentative evidence for a molecular bar.
From previous observations, UGC 2885 has not been reported to be a barred galaxy, and a preliminary analysis of the HST/WFC3 F814W images does not appear to show a stellar bar.

Galaxy metallicity is typically affected by star formation, inflows and outflows in the circumgalactic environment. A bar within the galaxy may contribute to the inflow and outflow processes. UGC~2885, being an isolated galaxy, is unlikely to be significantly affected by external inflow processes. The AGN suggested by \citet{holwerda2021predicting} may contribute to the outflow. It is uncertain how heavily AGN outflow affects the metallicity and star formation within the galaxy, as it is still unclear how the AGN is fueled.

Galaxies with strong bars usually have high metallicity values. Bars facilitate outer-disc to inner-disc inflow potential, which means that barred galaxies can process gas into stars more rapidly, thus increasing central metallicity \citep{vera2016effect}. Another possible effect of a bar on the galactic metallicity is the shape of the metallicity gradient. \citet{chen2023metallicity} showed that barred galaxies can flatten the radial metallicity profile. We observe this flattening for UGC~2885 in Figure \ref{fig:metallicityplotvalue}. 
Therefore, we understand that the presence of a molecular gas bar could explain some of the properties observed. Since this structure is not noticed in any other wavebands, we propose that its composition is mainly molecular and newly acquired by UGC~2885. Stronger evidence is needed to confirm this hypothesis, such as high-resolution sub-mm observations.

\section{Conclusions}\label{sec:conc}

We present a multiwavelength imaging analysis for UGC~2885, an extreme spiral in the nearby Universe with remarkably ordinary properties aside from its high molecular gas content. UGC~2885 has a typical brightness disc with signs of star formation along its arms, as well as a relatively low central activity. We compute both archival and new data over a wide range of wavelengths to provide an overview of several properties of this galaxy. In the following we summarise our findings:

\begin{enumerate}
    \item We used strong emission lines observed in SITELLE datacubes to calculate three different metallicity indicators  (N2O2, R23 and O3N2). The global metallicity is $Z \approx 9.0\pm 0.13$, which places UGC
    2885 in the high metallicity regime consistent with its large stellar mass. 
    \item There is little evidence for a strong metallicity gradient in UGC~2885 in any of the three metallicity indicators. Figure~\ref{fig:metallicityplotvalue} shows that the metallicity is uniform within $\pm0.3$~dex over a large radial extent in this galaxy.
    \item UGC~2885 aligns with the fundamental metallicity relation, indicating based on its stellar mass and star formation rate, a metallicity of $Z \approx 9.06$. 
    \item We find a mean global SFR of $1.63 \pm 0.72 \; \rm{M}_{\odot} \; \rm{yr}^{-1}$ derived from WISE W3 and W4 imaging. This is comparable to the only available calculated SFR in the literature for this object \citep[SFR = $2.47 \; \rm{M}_{\odot} \; \rm{yr}^{-1}$,][]{hunter2013star} assuming the same uncertainty range for their result. This represents a low star formation rate when we compare UGC~2885 to super spiral galaxies ($5 - 65 \; \rm{M}_{\odot} \; \rm{yr}^{-1}$), objects that are generally positioned above the main sequence of star forming galaxies in the nearby Universe. 
    \item The integrated molecular hydrogen mass (M$_{\rm H_{2}}$) is calculated to be $1.89 \pm 0.24 \times 10^{11} \; \rm{M}_{\odot}$. Galaxies at the higher end of stellar mass generally have much lower H$_{2}$ masses, as we expect the molecular gas to be consumed over the galaxy's lifetime by star formation, black hole accretion or external processes. This indicates that UGC~2885 is undergoing a quenching event, supported by its high depletion times based both on the molecular (t$_{\rm dep}~\rm{(H_2)} = 1.15 \pm 0.51 \times 10^{11}$~yr) and neutral hydrogen (t$_{\rm dep}~\rm{(HI)} = 2.29 \pm 1.04 \times 10^{10} $~yr) content.
    \item High t$_{\rm dep}$ is characteristic of galaxies below the star forming main sequence. Although UGC~2885 is located at the higher end of the distributions for stellar mass, its actual position in relation to other massive galaxies changes dramatically when we include the relationship between SFR and molecular mass. The position of galaxies in the MS is then determined by a combination of factors--- mainly the molecular gas fraction and the SFE \citep{saintonge2016molecular}--- as seen in Figures \ref{fig:depms} and \ref{fig:lisenfeldsfms}. 
    \item UGC~2885 has an extreme molecular-to-stellar mass ratio ($f_{\rm H_{2}} = -0.41 \pm 0.33$). Mechanisms that could add cold gas to this galaxy remain unknown as it is completely isolated without companions or a clear recent merging event.
    \item We suggest a barred structure in UGC~2885, mainly formed by molecular mass. The moment-0 and moment-2 maps show that UGC~2885's molecular mass is located in its central region, with a disturbed kinematics. This is the first study of the H$_{2}$ gas of UGC~2885, which could explain why no other previous work on this galaxy has pointed to the existence of a bar. The bar mechanisms that control the gas dynamics--- specifically inflows--- within the central regions galaxies are potential drivers of star formation quenching and high metallicity,  consistent with the observations in this study. 
\end{enumerate}

We conclude that UGC~2885 has produced stars through many cycles of star formation and has undergone no significant recent starburst or merging events. This might be responsible for the galaxy's current stage of evolution in terms of stellar content, population (metal richness) and star formation rate. The processes that fuel the large molecular gas reservoir remain uncertain. Possible directions for future work include spatially resolved analysis of star formation, gas content, gas and stellar kinematics, and morphological decomposition to infer more details about the galaxy's structure at all radii, from the putative bar to the spiral arms. Such an analysis would benefit from molecular gas and dust maps at higher spatial resolution than currently available.

\begin{acknowledgements}
The authors thank the anonymous referee for a helpful report.
We thank Dr. C. Rhea for helpful discussions and guidance in the use of LUCI, Dr. D. Hunter for kindly providing the WSRT 21-cm line observations, C. Robertson for helpful comments on the galaxy's spectra, and H. Christie for assistance with Figure~1.

Based on observations obtained with SITELLE, a joint project between Universit\'e Laval, ABB-Bomem, Universit\'e de Montr\'eal, and the Canada-France-Hawaii Telescope (CFHT) with funding support from the Canada Foundation for Innovation (CFI), the National Sciences and Engineering Research Council of Canada (NSERC), Fond de Recherche du Quebec - Nature et Technologies (FRQNT) and CFHT. The Canada-France-Hawaii Telescope is operated from the summit of Maunakea by the National Research Council of Canada, the Institut National des Sciences de l'Univers of the Centre National de la Recherche Scientifique of France, and the University of Hawaii. The observations at the Canada-France-Hawaii Telescope were performed with care and respect from the summit of Maunakea which is a significant cultural and historic site.

This work is based on observations carried out under project number 076-21 with the IRAM  30m telescope. IRAM is supported by INSU/CNRS (France), MPG (Germany) and IGN (Spain).

This research is based on observations made with the NASA/ESA Hubble Space Telescope obtained from the Space Telescope Science Institute, which is operated by the Association of Universities for Research in Astronomy, Inc., under NASA contract NAS 5–26555. These observations are associated with program(s), GO-15107 (PI B.W. Holwerda).

P. B. acknowledges support from a Natural Sciences and Engineering Research Council of Canada Discovery Grant.
S. K. gratefully acknowledges funding from the European Research Council (ERC) under the European Union's Horizon 2020 research and innovation programme (grant agreement No. 789410).
\end{acknowledgements}

\bibliographystyle{aa}
\bibliography{ugc2885}

\begin{thebibliography}{96}
\expandafter\ifx\csname natexlab\endcsname\relax\def\natexlab#1{#1}\fi

\bibitem[{{Alloin} {et~al.}(1979){Alloin}, {Collin-Souffrin}, {Joly}, \& {Vigroux}}]{1979A&A....78..200A}
{Alloin}, D., {Collin-Souffrin}, S., {Joly}, M., \& {Vigroux}, L. 1979, A\&A, 78, 200

\bibitem[{Arimoto {et~al.}(1996)Arimoto, Sofue, \& Tsujimoto}]{arimoto1996co}
Arimoto, N., Sofue, Y., \& Tsujimoto, T. 1996, PASJ, 48, 275

\bibitem[{Asplund {et~al.}(2009)Asplund, Grevesse, Sauval, \& Scott}]{asplund2009chemical}
Asplund, M., Grevesse, N., Sauval, A.~J., \& Scott, P. 2009, ARAA, 47, 481

\bibitem[{Baars {et~al.}(1987)Baars, Hooghoudt, Mezger, \& de~Jonge}]{baars1987iram}
Baars, J., Hooghoudt, B., Mezger, P., \& de~Jonge, M. 1987, A\&A, 175, 319

\bibitem[{Baker {et~al.}(2023)Baker, Maiolino, Belfiore, Bluck, Curti, Wylezalek, Bertemes, Bothwell, Lin, Thorp, {et~al.}}]{baker2023molecular}
Baker, W.~M., Maiolino, R., Belfiore, F., {et~al.} 2023, MNRAS, 518, 4767

\bibitem[{Barbary(2016)}]{barbary2016extinction}
Barbary, K. 2016, Zenodo

\bibitem[{Bolatto {et~al.}(2013)Bolatto, Wolfire, \& Leroy}]{bolatto2013co}
Bolatto, A.~D., Wolfire, M., \& Leroy, A.~K. 2013, ARAA, 51, 207

\bibitem[{Brinchmann {et~al.}(2004)Brinchmann, Charlot, White, Tremonti, Kauffmann, Heckman, \& Brinkmann}]{brinchmann2004physical}
Brinchmann, J., Charlot, S., White, S.~D., {et~al.} 2004, MNRAS, 351, 1151

\bibitem[{Brown {et~al.}(2017)Brown, Moustakas, Kennicutt, Bonne, Intema, De~Gasperin, Boquien, Jarrett, Cluver, Smith, {et~al.}}]{brown2017calibration}
Brown, M.~J., Moustakas, J., Kennicutt, R.~C., {et~al.} 2017, ApJ, 847, 136

\bibitem[{Brown {et~al.}(2014)Brown, Jarrett, \& Cluver}]{brown2014recalibrating}
Brown, M. J.~I., Jarrett, T., \& Cluver, M.~E. 2014, PASA, 31, e049

\bibitem[{Canzian {et~al.}(1993)Canzian, Allen, \& Tilanus}]{canzian1993spiral}
Canzian, B., Allen, R., \& Tilanus, R. 1993, ApJ, 406, 457

\bibitem[{{Cardelli} {et~al.}(1989){Cardelli}, {Clayton}, \& {Mathis}}]{1989ApJ...345..245C}
{Cardelli}, J.~A., {Clayton}, G.~C., \& {Mathis}, J.~S. 1989, ApJ, 345

\bibitem[{Chen {et~al.}(2023)Chen, Grasha, Battisti, Kewley, Madore, Seibert, Rich, \& Beaton}]{chen2023metallicity}
Chen, Q.-H., Grasha, K., Battisti, A.~J., {et~al.} 2023, MNRAS, 519, 4801

\bibitem[{Chiang(2023)}]{chiang2023corrected}
Chiang, Y.-K. 2023, ApJ, 958, 118

\bibitem[{Cluver {et~al.}(2017)Cluver, Jarrett, Dale, Smith, August, \& Brown}]{cluver2017calibrating}
Cluver, M., Jarrett, T.~H., Dale, D., {et~al.} 2017, ApJ, 850, 68

\bibitem[{Cluver {et~al.}(2014)Cluver, Jarrett, Hopkins, Driver, Liske, Gunawardhana, Taylor, Robotham, Alpaslan, Baldry, {et~al.}}]{cluver2014galaxy}
Cluver, M.~E., Jarrett, T.~H., Hopkins, A.~M., {et~al.} 2014, ApJ, 782, 90

\bibitem[{Cutri {et~al.}(2012)Cutri, Wright, Conrow, Bauer, Benford, Brandenburg, Dailey, Eisenhardt, Evans, Fajardo-Acosta, {et~al.}}]{cutri2012explanatory}
Cutri, R., Wright, E., Conrow, T., {et~al.} 2012, Explanatory Supplement to the WISE All-Sky Data Release Products, 1

\bibitem[{Cutri {et~al.}(2021)Cutri, Wright, Conrow, Fowler, Eisenhardt, Grillmair, Kirkpatrick, Masci, McCallon, Wheelock, {et~al.}}]{cutri2021vizier}
Cutri, R.~e., Wright, E., Conrow, T., {et~al.} 2021, VizieR Online Data Catalog, II

\bibitem[{De~Vaucouleurs {et~al.}(2013)De~Vaucouleurs, de~Vaucouleurs, Harold~Jr, Buta, Paturel, Fouque, {et~al.}}]{de2013third}
De~Vaucouleurs, G., de~Vaucouleurs, A., Harold~Jr, G., {et~al.} 2013, Third Reference Catalogue of Bright Galaxies: Volume III, Vol.~3 (Springer Science \& Business Media)

\bibitem[{Di~Teodoro {et~al.}(2023)Di~Teodoro, Posti, Fall, Ogle, Jarrett, Appleton, Cluver, Haynes, \& Lisenfeld}]{di2023dark}
Di~Teodoro, E.~M., Posti, L., Fall, S.~M., {et~al.} 2023, MNRAS, 518, 6340

\bibitem[{Dietrich {et~al.}(2018)Dietrich, Weiner, Ashby, Hayward, Mart{\'\i}nez-Galarza, Ramos~Padilla, Rosenthal, Smith, Willner, \& Zezas}]{dietrich2018agn}
Dietrich, J., Weiner, A.~S., Ashby, M.~L., {et~al.} 2018, MNRAS, 480, 3562

\bibitem[{{Dopita} \& {Evans}(1986)}]{1986ApJ...307..431D}
{Dopita}, M.~A. \& {Evans}, I.~N. 1986, ApJ, 307, 431

\bibitem[{Doyle \& Drinkwater(2006)}]{doyle2006effect}
Doyle, M.~T. \& Drinkwater, M.~J. 2006, MNRAS, 372, 977

\bibitem[{Drissen {et~al.}(2019)Drissen, Martin, Rousseau-Nepton, Robert, Martin, Baril, Prunet, Joncas, Thibault, Brousseau, Mandar, Grandmont, Yee, \& Simard}]{drissen_sitelle_2019}
Drissen, L., Martin, T., Rousseau-Nepton, L., {et~al.} 2019, MNRAS, 485, 3930

\bibitem[{Du {et~al.}(2023)Du, Cheng, Du, Du, \& Wu}]{du2023star}
Du, W., Cheng, C., Du, P., Du, L., \& Wu, H. 2023, ApJ, 959, 105

\bibitem[{{Edmunds} \& {Pagel}(1984)}]{1984MNRAS.211..507E}
{Edmunds}, M.~G. \& {Pagel}, B.~E.~J. 1984, MNRAS, 211, 507

\bibitem[{Emonts {et~al.}(2014)Emonts, Norris, Feain, Mao, Ekers, Miley, Seymour, R{\"o}ttgering, Villar-Martin, Sadler, {et~al.}}]{emonts2014co}
Emonts, B., Norris, R.~P., Feain, I., {et~al.} 2014, MNRAS, 438, 2898

\bibitem[{Falco {et~al.}(1999)Falco, Kurtz, Geller, Huchra, Peters, Berlind, Mink, Tokarz, \& Elwell}]{falco1999updated}
Falco, E.~E., Kurtz, M.~J., Geller, M.~J., {et~al.} 1999, PASP, 111, 438

\bibitem[{Fitzpatrick \& Massa(2007)}]{fitzpatrick2007analysis}
Fitzpatrick, E. \& Massa, D. 2007, ApJ, 663, 320

\bibitem[{{French}(2021)}]{french2021}
{French}, K.~D. 2021, \pasp, 133, 072001

\bibitem[{Gao {et~al.}(2020)Gao, Wang, Pearson, Gordon, Holwerda, Hopkins, Brown, Bland-Hawthorn, \& Owers}]{gao2020mergers}
Gao, F., Wang, L., Pearson, W., {et~al.} 2020, A\&A, 637, A94

\bibitem[{Ginsburg {et~al.}(2015)Ginsburg, Robitaille, Beaumont, Rosolowsky, Leroy, Brogan, Hunter, Teuben, \& Brisbin}]{ginsburg2015radio}
Ginsburg, A., Robitaille, T., Beaumont, C., {et~al.} 2015, Revolution in Astronomy with ALMA: The Third Year, 499, 363

\bibitem[{Harris {et~al.}(2010)Harris, Baker, Zonak, Sharon, Genzel, Rauch, Watts, \& Creager}]{harris2010co}
Harris, A., Baker, A., Zonak, S., {et~al.} 2010, ApJ, 723, 1139

\bibitem[{{Hill} {et~al.}(2008)}]{2008SPIE.7014E..70H}
{Hill}, G.~J. {et~al.} 2008, in Society of Photo-Optical Instrumentation Engineers (SPIE) Conference Series, Vol. 7014, Ground-based and Airborne Instrumentation for Astronomy II, ed. I.~S. {McLean} \& M.~M. {Casali}, 701470

\bibitem[{Hirashita(2023)}]{hirashita2023effects}
Hirashita, H. 2023, MNRAS, 522, 4612

\bibitem[{{Ho} {et~al.}(2015){Ho}, {Kudritzki}, {Kewley}, {Zahid}, {Dopita}, {Bresolin}, \& {Rupke}}]{ho_metallicity_2015}
{Ho}, I.~T., {Kudritzki}, R.-P., {Kewley}, L.~J., {et~al.} 2015, MNRAS, 448, 2030

\bibitem[{Holwerda {et~al.}(2021)Holwerda, Wu, Keel, Young, Mullins, Hinz, Ford, Barmby, Chandar, Bailin, {et~al.}}]{holwerda2021predicting}
Holwerda, B.~W., Wu, J.~F., Keel, W.~C., {et~al.} 2021, ApJ, 914, 142

\bibitem[{Hopkins {et~al.}(2013)Hopkins, Driver, Brough, Owers, Bauer, Gunawardhana, Cluver, Colless, Foster, Lara-L{\'o}pez, {et~al.}}]{hopkins2013galaxy}
Hopkins, A.~M., Driver, S.~P., Brough, S., {et~al.} 2013, MNRAS, 430, 2047

\bibitem[{Hopkins {et~al.}(2012)Hopkins, Quataert, \& Murray}]{hopkins2012structure}
Hopkins, P.~F., Quataert, E., \& Murray, N. 2012, MNRAS, 421, 3488

\bibitem[{Hunter {et~al.}(2013)Hunter, Elmegreen, Rubin, Ashburn, Wright, J{\'o}zsa, \& Struve}]{hunter2013star}
Hunter, D.~A., Elmegreen, B.~G., Rubin, V.~C., {et~al.} 2013, AJ, 146, 92

\bibitem[{Jarrett {et~al.}(2011)Jarrett, Cohen, Masci, Wright, Stern, Benford, Blain, Carey, Cutri, Eisenhardt, {et~al.}}]{jarrett2011spitzer}
Jarrett, T., Cohen, M., Masci, F., {et~al.} 2011, ApJ, 735, 112

\bibitem[{Jarrett {et~al.}(2012)Jarrett, Masci, Tsai, Petty, Cluver, Assef, Benford, Blain, Bridge, Donoso, {et~al.}}]{jarrett2012extending}
Jarrett, T., Masci, F., Tsai, C., {et~al.} 2012, AJ, 145, 6

\bibitem[{{Jarrett} {et~al.}(2023){Jarrett}, {Cluver}, {Taylor}, {Bellstedt}, {Robotham}, \& {Yao}}]{jarrett2023new}
{Jarrett}, T.~H., {Cluver}, M.~E., {Taylor}, E.~N., {et~al.} 2023, \apj, 946, 95

\bibitem[{Jim{\'e}nez-Donaire {et~al.}(2019)Jim{\'e}nez-Donaire, Bigiel, Leroy, Usero, Cormier, Puschnig, Gallagher, Kepley, Bolatto, Garc{\'\i}a-Burillo, {et~al.}}]{jimenez2019empire}
Jim{\'e}nez-Donaire, M.~J., Bigiel, F., Leroy, A., {et~al.} 2019, ApJ, 880, 127

\bibitem[{Joye \& Mandel(2003)}]{joye2003new}
Joye, W.~A. \& Mandel, E. 2003, in Astronomical data analysis software and systems XII, Vol. 295, 489

\bibitem[{Kennicutt {et~al.}(2011)Kennicutt, Calzetti, Aniano, Appleton, Armus, Beirao, Bolatto, Brandl, Crocker, Croxall, {et~al.}}]{kennicutt2011kingfish}
Kennicutt, R., Calzetti, D., Aniano, G., {et~al.} 2011, PASP, 123, 1347

\bibitem[{Kennicutt {et~al.}(2003)Kennicutt, Armus, Bendo, Calzetti, Dale, Draine, Engelbracht, Gordon, Grauer, Helou, {et~al.}}]{kennicutt2003sings}
Kennicutt, R.~C., Armus, L., Bendo, G., {et~al.} 2003, PASP, 115, 928

\bibitem[{{Kennicutt} \& {Evans}(2012)}]{ke2012}
{Kennicutt}, R.~C. \& {Evans}, N.~J. 2012, \araa, 50, 531

\bibitem[{Kere{\v{s}} {et~al.}(2005)Kere{\v{s}}, Katz, Weinberg, \& Dav{\'e}}]{kerevs2005galaxies}
Kere{\v{s}}, D., Katz, N., Weinberg, D.~H., \& Dav{\'e}, R. 2005, MNRAS, 363, 2

\bibitem[{Kettlety {et~al.}(2018)Kettlety, Hesling, Phillipps, Bremer, Cluver, Taylor, Bland-Hawthorn, Brough, De~Propris, Driver, {et~al.}}]{kettlety2018galaxy}
Kettlety, T., Hesling, J., Phillipps, S., {et~al.} 2018, MNRAS, 473, 776

\bibitem[{{Kewley} \& {Dopita}(2002)}]{2002ApJS..142...35K}
{Kewley}, L.~J. \& {Dopita}, M.~A. 2002, ApJS, 142, 35

\bibitem[{{Kewley} {et~al.}(2006){Kewley}, {Groves}, {Kauffmann}, \& {Heckman}}]{kewley2006}
{Kewley}, L.~J., {Groves}, B., {Kauffmann}, G., \& {Heckman}, T. 2006, \mnras, 372, 961

\bibitem[{Kim \& Ostriker(2007)}]{kim2007gravitational}
Kim, W.-T. \& Ostriker, E.~C. 2007, ApJ, 660, 1232

\bibitem[{{Kobulnicky} \& {Kewley}(2004)}]{2004ApJ...617..240K}
{Kobulnicky}, H.~A. \& {Kewley}, L.~J. 2004, ApJ, 617, 240

\bibitem[{Lacey \& Cole(1993)}]{lacey1993merger}
Lacey, C. \& Cole, S. 1993, MNRAS, 262, 627

\bibitem[{Leroy {et~al.}(2019)Leroy, Sandstrom, Lang, Lewis, Salim, Behrens, Chastenet, Chiang, Gallagher, Kessler, {et~al.}}]{leroy2019z}
Leroy, A.~K., Sandstrom, K.~M., Lang, D., {et~al.} 2019, ApJS, 244, 24

\bibitem[{Leroy {et~al.}(2008)Leroy, Walter, Brinks, Bigiel, De~Blok, Madore, \& Thornley}]{leroy2008star}
Leroy, A.~K., Walter, F., Brinks, E., {et~al.} 2008, AJ, 136, 2782

\bibitem[{Lewis(1985)}]{lewis1985hi}
Lewis, B. 1985, ApJ, 292, 451

\bibitem[{Lin {et~al.}(2019)Lin, Pan, Ellison, Belfiore, Shi, S{\'a}nchez, Hsieh, Rowlands, Ramya, Thorp, {et~al.}}]{lin2019almaquest}
Lin, L., Pan, H.-A., Ellison, S.~L., {et~al.} 2019, ApJL, 884, L33

\bibitem[{{Lisenfeld} {et~al.}(2023){Lisenfeld}, {Ogle}, {Appleton}, {Jarrett}, \& {Moncada-Cuadri}}]{lisenfeld2023molecular}
{Lisenfeld}, U., {Ogle}, P.~M., {Appleton}, P.~N., {Jarrett}, T.~H., \& {Moncada-Cuadri}, B.~M. 2023, A\&A, 673, A87

\bibitem[{{Looser} {et~al.}(2024){Looser}, {D'Eugenio}, {Piotrowska}, {Belfiore}, {Maiolino}, {Cappellari}, {Baker}, \& {Tacchella}}]{looser2024stellar}
{Looser}, T.~J., {D'Eugenio}, F., {Piotrowska}, J.~M., {et~al.} 2024, \mnras, 532, 2832

\bibitem[{Mannucci {et~al.}(2010)Mannucci, Cresci, Maiolino, Marconi, \& Gnerucci}]{Mannucci_2010}
Mannucci, F., Cresci, G., Maiolino, R., Marconi, A., \& Gnerucci, A. 2010, MNRAS, 408, 2115

\bibitem[{{Martin} {et~al.}(2012){Martin}, {Drissen}, \& {Joncas}}]{martin2012}
{Martin}, T., {Drissen}, L., \& {Joncas}, G. 2012, in Society of Photo-Optical Instrumentation Engineers (SPIE) Conference Series, Vol. 8451, Software and Cyberinfrastructure for Astronomy II, ed. N.~M. {Radziwill} \& G.~{Chiozzi}, 84513K

\bibitem[{{McCall} {et~al.}(1985){McCall}, {Rybski}, \& {Shields}}]{1985ApJS...57....1M}
{McCall}, M.~L., {Rybski}, P.~M., \& {Shields}, G.~A. 1985, ApJS, 57, 1

\bibitem[{Meidt {et~al.}(2014)Meidt, Schinnerer, Van~de Ven, Zaritsky, Peletier, Knapen, Sheth, Regan, Querejeta, Mu{\~n}oz-Mateos, {et~al.}}]{meidt2014reconstructing}
Meidt, S.~E., Schinnerer, E., Van~de Ven, G., {et~al.} 2014, ApJ, 788, 144

\bibitem[{{Moustakas} {et~al.}(2023){Moustakas}, {Lang}, {Dey}, {Juneau}, {Meisner}, {Myers}, {Schlafly}, {Schlegel}, {Valdes}, {Weaver}, \& {Zhou}}]{sga2023}
{Moustakas}, J., {Lang}, D., {Dey}, A., {et~al.} 2023, \apjs, 269, 3

\bibitem[{Ogle {et~al.}(2016)Ogle, Lanz, Nader, \& Helou}]{ogle2016superluminous}
Ogle, P.~M., Lanz, L., Nader, C., \& Helou, G. 2016, ApJ, 817, 109

\bibitem[{Paalvast \& Brinchmann(2017)}]{Paalvast_2017}
Paalvast, M. \& Brinchmann, J. 2017, MNRAS, 470, 1612

\bibitem[{{Pagel} {et~al.}(1979){Pagel}, {Edmunds}, {Blackwell}, {Chun}, \& {Smith}}]{1979MNRAS.189...95P}
{Pagel}, B.~E.~J., {Edmunds}, M.~G., {Blackwell}, D.~E., {Chun}, M.~S., \& {Smith}, G. 1979, MNRAS, 189, 95

\bibitem[{Parkash {et~al.}(2018)Parkash, Brown, Jarrett, \& Bonne}]{parkash2018relationships}
Parkash, V., Brown, M.~J., Jarrett, T., \& Bonne, N.~J. 2018, ApJ, 864, 40

\bibitem[{Peng {et~al.}(2015)Peng, Maiolino, \& Cochrane}]{peng2015strangulation}
Peng, Y., Maiolino, R., \& Cochrane, R. 2015, Nature, 521, 192

\bibitem[{{Pettini} \& {Pagel}(2004)}]{2004MNRAS.348L..59P}
{Pettini}, M. \& {Pagel}, B. E.~J. 2004, MNRAS, 348, L59

\bibitem[{Ray {et~al.}(2024)Ray, Dhiwar, Bagchi, \& Pandge}]{ray2024probing}
Ray, S., Dhiwar, S., Bagchi, J., \& Pandge, M. 2024, MNRAS, 527, 9999

\bibitem[{Rhea {et~al.}(2021{\natexlab{a}})Rhea, Hlavacek-Larrondo, Rousseau-Nepton, Vigneron, \& Guité}]{rhea_luci_2021}
Rhea, C., Hlavacek-Larrondo, J., Rousseau-Nepton, L., Vigneron, B., \& Guité, L.-S. 2021{\natexlab{a}}, RNAAS, 208

\bibitem[{Rhea {et~al.}(2021{\natexlab{b}})Rhea, Rousseau-Nepton, Covington, Alcorn, Vigneron, Hlavacek-Larrondo, \& Guité}]{Carter_2021}
Rhea, C.~L., Rousseau-Nepton, L., Covington, J., {et~al.} 2021{\natexlab{b}}, crhea93/LUCI: Luci Updates (v1.1). Zenodo.

\bibitem[{Roelfsema \& Allen(1985)}]{roelfsema1985radio}
Roelfsema, P. \& Allen, R. 1985, A\&A, 146, 213

\bibitem[{Romanishin(1983)}]{romanishin1983very}
Romanishin, W. 1983, MNRAS, 204, 909

\bibitem[{Rubin {et~al.}(1980)Rubin, Ford~Jr, \& Thonnard}]{rubin1980rotational}
Rubin, V.~C., Ford~Jr, W.~K., \& Thonnard, N. 1980, ApJ, 238, 471

\bibitem[{Saburova {et~al.}(2021)Saburova, Chilingarian, Kasparova, Sil’chenko, Grishin, Katkov, \& Uklein}]{saburova2021observational}
Saburova, A.~S., Chilingarian, I.~V., Kasparova, A.~V., {et~al.} 2021, MNRAS, 503, 830

\bibitem[{Saintonge {et~al.}(2016)Saintonge, Catinella, Cortese, Genzel, Giovanelli, Haynes, Janowiecki, Kramer, Lutz, Schiminovich, {et~al.}}]{saintonge2016molecular}
Saintonge, A., Catinella, B., Cortese, L., {et~al.} 2016, MNRAS, 462, 1749

\bibitem[{Saintonge {et~al.}(2017)Saintonge, Catinella, Tacconi, Kauffmann, Genzel, Cortese, Dav{\'e}, Fletcher, Graci{\'a}-Carpio, Kramer, {et~al.}}]{saintonge2017xcold}
Saintonge, A., Catinella, B., Tacconi, L.~J., {et~al.} 2017, ApJS, 233, 22

\bibitem[{{Salim} {et~al.}(2016){Salim}, {Lee}, {Janowiecki}, {da Cunha}, {Dickinson}, {Boquien}, {Burgarella}, {Salzer}, \& {Charlot}}]{GSWLC2016}
{Salim}, S., {Lee}, J.~C., {Janowiecki}, S., {et~al.} 2016, \apjs, 227, 2

\bibitem[{Sandstrom {et~al.}(2010)Sandstrom, Bolatto, Draine, Bot, \& Stanimirovi{\'c}}]{sandstrom2010spitzer}
Sandstrom, K.~M., Bolatto, A.~D., Draine, B., Bot, C., \& Stanimirovi{\'c}, S. 2010, ApJ, 715, 701

\bibitem[{Sandstrom {et~al.}(2024)Sandstrom, Chastenet, Bolatto, Koch, Leroy, Teng, Williams, {et~al.}}]{sandstrom2024resolved}
Sandstrom, K.~M., Chastenet, J., Bolatto, A.~D., {et~al.} 2024, ApJ, 964

\bibitem[{Sandstrom {et~al.}(2013)Sandstrom, Leroy, Walter, Bolatto, Croxall, Draine, Wilson, Wolfire, Calzetti, Kennicutt, {et~al.}}]{sandstrom2013co}
Sandstrom, K.~M., Leroy, A., Walter, F., {et~al.} 2013, ApJ, 777, 5

\bibitem[{Scudder {et~al.}(2021)Scudder, Ellison, El Meddah El Idrissi, \& Poetrodjojo}]{scudder2021}
Scudder, J.~M., Ellison, S.~L., El Meddah El Idrissi, L., \& Poetrodjojo, H. 2021, MNRAS, 507, 2468

\bibitem[{Sharda {et~al.}(2021)Sharda, Krumholz, Wisnioski, Forbes, Federrath, \& Acharyya}]{Sharda_2021}
Sharda, P., Krumholz, M.~R., Wisnioski, E., {et~al.} 2021, MNRAS, 502

\bibitem[{Skrutskie {et~al.}(2006)Skrutskie, Cutri, Stiening, Weinberg, Schneider, Carpenter, Beichman, Capps, Chester, Elias, {et~al.}}]{skrutskie2006two}
Skrutskie, M., Cutri, R., Stiening, R., {et~al.} 2006, AJ, 131, 1163

\bibitem[{{Snyder} {et~al.}(2015){Snyder}, {Torrey}, {Lotz}, {Genel}, {McBride}, {Vogelsberger}, {Pillepich}, {Nelson}, {Sales}, {Sijacki}, {Hernquist}, \& {Springel}}]{snyder2015}
{Snyder}, G.~F., {Torrey}, P., {Lotz}, J.~M., {et~al.} 2015, MNRAS, 454, 1886

\bibitem[{Strong \& Mattox(1996)}]{strong1996gradient}
Strong, A. \& Mattox, J. 1996, A\&A, 308, L21

\bibitem[{Sun {et~al.}(2023)Sun, Leroy, Ostriker, Meidt, Rosolowsky, Schinnerer, Wilson, Utomo, Belfiore, Blanc, {et~al.}}]{sun2023star}
Sun, J., Leroy, A.~K., Ostriker, E.~C., {et~al.} 2023, ApJL, 945, L19

\bibitem[{Tissera {et~al.}(2021)Tissera, Rosas-Guevara, Sillero, Pedrosa, Theuns, \& Bignone}]{Tissera_2021}
Tissera, P.~B., Rosas-Guevara, Y., Sillero, E., {et~al.} 2021, MNRAS, 511

\bibitem[{Vera {et~al.}(2016)Vera, Alonso, \& Coldwell}]{vera2016effect}
Vera, M., Alonso, S., \& Coldwell, G. 2016, A\&A, 595, A63

\bibitem[{{Vogelsberger} {et~al.}(2014){Vogelsberger}, {Genel}, {Springel}, {Torrey}, {Sijacki}, {Xu}, {Snyder}, {Nelson}, \& {Hernquist}}]{vogels2014}
{Vogelsberger}, M., {Genel}, S., {Springel}, V., {et~al.} 2014, MNRAS, 444, 1518

\bibitem[{Wright {et~al.}(2010)Wright, Eisenhardt, Mainzer, Ressler, Cutri, Jarrett, Kirkpatrick, Padgett, McMillan, Skrutskie, {et~al.}}]{wright2010wide}
Wright, E.~L., Eisenhardt, P.~R., Mainzer, A.~K., {et~al.} 2010, AJ, 140, 1868

\bibitem[{Zahid {et~al.}(2014)Zahid, Dima, Kudritzki, Kewley, Geller, Hwang, Silverman, \& Kashino}]{Zahid_2014}
Zahid, H.~J., Dima, G.~I., Kudritzki, R.-P., {et~al.} 2014, ApJ, 791, 130

\end{thebibliography}

\begin{appendix}
\section{Nuclear spectra and line ratio diagrams}
\label{sec:apdx}

\begin{figure}
    \centering
    \includegraphics[width=\columnwidth]{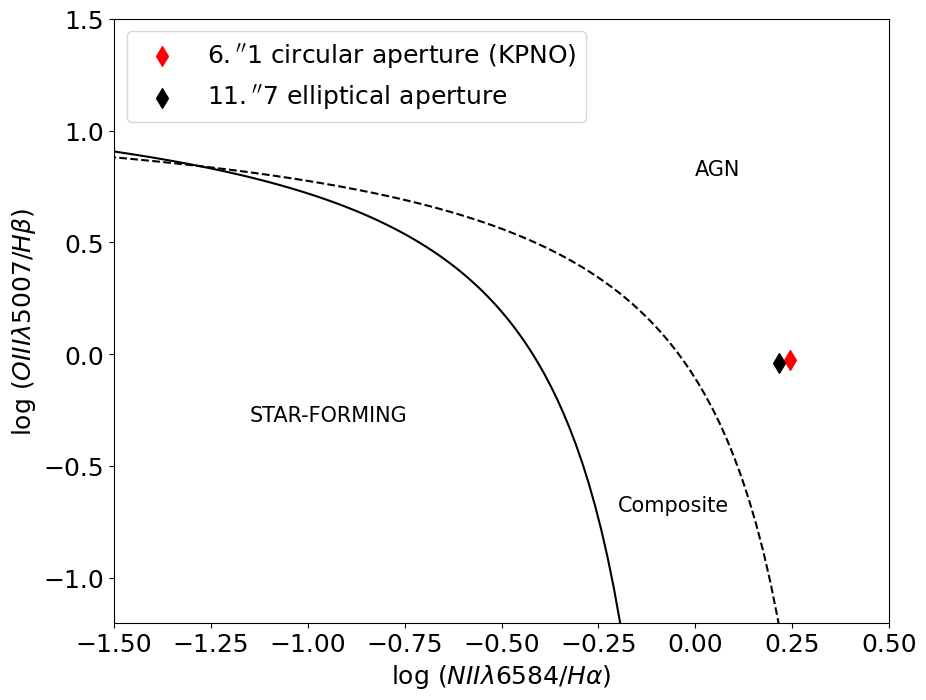}
    \includegraphics[width=\columnwidth]{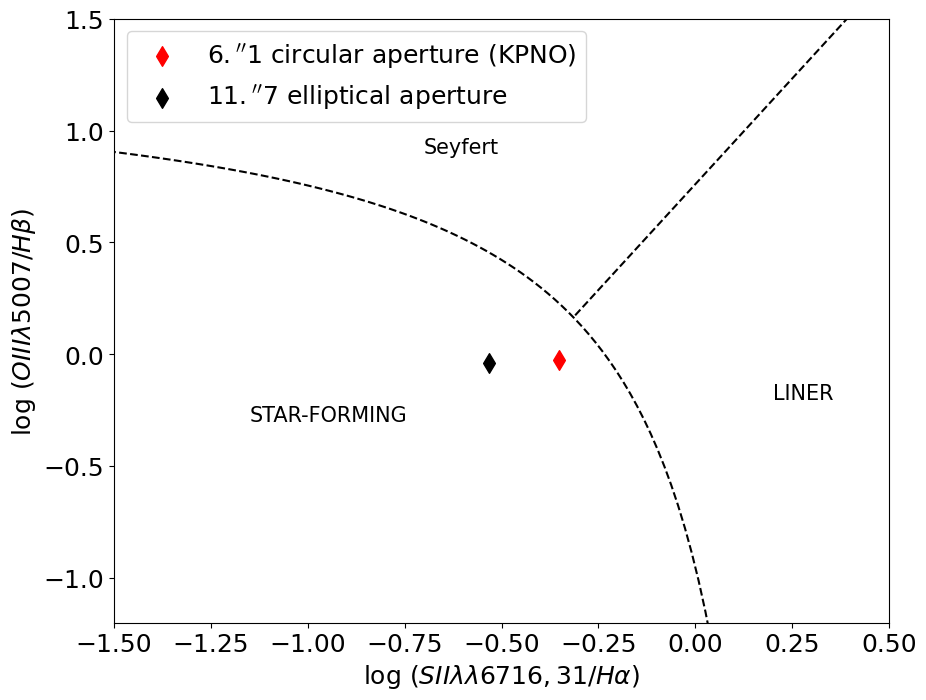}
    \caption{Emission line-ratio (``BPT'') diagrams showing the position of the UGC~2885 nucleus. Spectra were extracted in two apertures, with radii 6\farcs1 and 11\farcs7 to match previous observations.}
    \label{fig:bpt}
\end{figure}

As described in Section~\ref{sec:sitelle_data}, the major purpose for obtaining SITELLE observations was to characterise UGC 2885's spatially-resolved metallicity. 
Since AGN activity has implications for the galaxy's evolution,
 these data can also be used to extract nuclear emission line ratios for comparison with the measurements reported by \citet{holwerda2021predicting}.
Using the line flux maps described in Section~\ref{sec:sitelle_data}, we computed emission line ratios over two apertures of
radii 6\farcs1 and 11\farcs7 to match previous observations.
Figure~\ref{fig:bpt} shows the measurements in line ratio diagrams, with boundaries between object types as defined by \citet{kewley2006}.
The \ion{O}{III}/H$\beta$ versus \ion{N}{II}/H$\alpha$ diagram (top panel) shows line ratios for UGC 2885 that fall  in the AGN region, while the \ion{O}{III}/H$\beta$ versus \ion{S}{II}/H$\alpha$ diagram (bottom panel) shows ratios that fall in the `star-forming' region near the border with `Seyfert' and `LINER' classifications.
The \ion{S}{II} lines in the SITELLE spectrum are relatively weak and the \ion{S}{II}/H$\alpha$ ratio more uncertain than \ion{O}{III}/H$\beta$ or \ion{N}{II}/H$\alpha$.
These results are consistent with those of \citet{holwerda2021predicting}.

\end{appendix}

\end{document}